\documentclass[12pt,a4paper,notitlepage,fleqn]{article}

\usepackage{setspace}
\usepackage[margin=1in]{geometry}
\usepackage[symbol]{footmisc}

\usepackage[T1]{fontenc}
\usepackage[utf8]{inputenc}
\usepackage{authblk}

\makeatletter

\newcommand{\fboxsubsec}[1]{
	\begin{flushleft}
		#1
	\end{flushleft}
	}
\renewcommand{\subsection}{\@startsection{subsection}{2}{0pt}
	{1ex}
	{0.5ex}
	{\reset@font\it\fboxsubsec}
	}
\makeatother

\title{Measuring the Time-Varying Market Efficiency\\
in the Prewar and Wartime Japanese Stock Market, 1924--1943}%

\author{Kenichi Hirayama$^{a}$ \ and \ Akihiko Noda$^{b,c}$\thanks{\scriptsize Corresponding Author. E-mail: anoda@meiji.ac.jp, Tel. +81-3-3296-2265, Fax. +81-3-3296-4347.}

{\scriptsize ${}^{a}$ \it Tokio Marine Asset Management Co., Ltd., 18th Floor Tekko Building, 1-8-2 Marunouchi, Tokyo 100-0005, Japan} 

{\scriptsize ${}^{b}$ \it School of Commerce, Meiji University, 1-1 Kanda-Surugadai, Chiyoda-ku, Tokyo 101-8301, Japan}

{\scriptsize ${}^{c}$ \it Keio Economic Observatory, Keio University, 2-15-45 Mita, Minato-ku, Tokyo 108-8345, Japan}}

\date{Forthcoming in {\it Asia-Pacific Economic History Review}\\
(accepted at May 15, 2024)}

\renewcommand\thefootnote{\arabic{footnote}}

\usepackage{graphicx}

\usepackage[sort]{natbib}%
\usepackage{amsmath,amssymb}%
\usepackage{ascmac}%
\usepackage{multirow}%
\usepackage{lscape}%
\usepackage{subfigmat}

\usepackage{pifont}%
\usepackage{arydshln}%
\usepackage[format=hang]{caption}
\usepackage[all]{xy}

\PassOptionsToPackage{hyphens}{url}
\usepackage{color}
\usepackage[bookmarkstype=toc,
	    colorlinks=true, 
	    pdfborder={0 0 1}, 
	    bookmarks=true, 
	    bookmarksnumbered=true, 
	    citecolor=blue]{hyperref}

\bibpunct{(}{)}{;}{a}{}{,}

\def\hsymbu#1{\smash{\lower1.7ex\hbox{\huge$#1$}}}

\def\ve #1{{\mbox{\boldmath $#1$}}}

\newcommand{\citetapos}[1]{\citeauthor{#1}'s \citeyearpar{#1}}
\newcommand{\citeapos}[2]{\citeauthor{#1}'s (\citeyear{#2})}
\newcommand{\ex}{{\mathbb{E}}}

\def\ve #1{{\mbox{\boldmath $#1$}}}

\begin{document}

\begin{titlepage}

\renewcommand{\thepage}{}
\renewcommand{\thefootnote}{\fnsymbol{footnote}}

\maketitle

\vspace{-10mm}

\noindent
\hrulefill

\noindent
{\bfseries Abstract:} This study explores the time-varying structure of market efficiency in the prewar and wartime Japanese stock market using a new market capitalization-weighted stock price index, the equity performance index. We examine whether the adaptive market hypothesis (AMH) is supported in that era. First, we find that the degree of market efficiency in the prewar and wartime Japanese stock market varies over time and with major historical events. This implies that the AMH is supported in this market. Second, we find that the variation in market efficiency observed in this study is significantly different from that in previous studies because of whether the price index is capitalization weighted. Finally, as government intervention in the market intensified throughout the 1930s, market efficiency declined as the war risk premium rose, especially from the time when the Pacific War became inevitable.\\



\noindent
\hrulefill

\end{titlepage}

\bibliographystyle{asa}


\section{Introduction}\label{prewar_stock_sec1}
Economists have shown interest in assessing whether \citetapos{fama1970ecm} efficient market hypothesis (EMH), that the market efficiently reflects all relevant information in determining the prices, is applicable to the stock market.\footnote{\citet{roberts1967scp} categorizes the three types of the EMH by differences in the available information sets: the weak sense (only the history of prices or returns themselves), the semi-strong sense (also includes publicly available information), and the strong sense (also includes private information).} However, there is no consensus on its applicability to stock markets, which is why the hypothesis remains controversial, as described by \citet{malkiel2004rwd} and \citet{shiller2005ies}. As such, \citet{lo2004amh} proposes the adaptive market hypothesis (AMH) as an alternative to the EMH. The AMH asserts that the market efficiency can vary over time due to changing market conditions (e.g., bubbles, crashes, and crises) and institutional factors. An essential implication of the AMH is that the stock prices do not always and fully reflect all relevant information in the market, and that it is not realistic to enquire whether the market is completely efficient, as per the EMH. Indeed, \citet{lo2004amh} shows that market efficiency has changed cyclically over time, with surprising evidence that the U.S. stock market was more efficient in the 1950s than in the early 1990s.

In an overview of the AMH literature, \citet{lim2011esm} show that various methods, time-varying parameter models, and conventional statistical inferences with rolling windows have been employed to explore whether stock market efficiency changes over time. Previous studies also detect the periods when the market has been inefficient to explore whether business cycles, bubbles, and economic crises in financial markets have specific dynamics and clarify whether their evolution determines such market dynamics, as mentioned by \citet{lo2005rem}. Studies such as \citet{ito2009mdt}, \citet{ito2014ism,ito2016eme}, \citet{kim2011srp}, \citet{lim2013aus}, and \cite{noda2016amh} affirm that stock markets evolve over time in accordance with the changes in market conditions. However, there are few studies on stock market efficiency from a historical perspective, owing to the low availability of prewar and wartime stock market data (except for the U.S. and Japan). 

Two major long-run datasets exist for the prewar and wartime U.S. stock market: the Dow Jones Industrial Average index and the S\&P composite index. \citet{choudhry2010ww2} investigates endogenous structural breaks in the U.S. stock market using the daily Dow Jones Industrial Average index during World War I\hspace{-.1em}I (WWI\hspace{-.1em}I). He finds that the market's breakpoints are consistent with major historical events during the war and that these events are correctly reflected in stock returns from the perspective of semi-strong market efficiency. \citet{kim2011srp} applies the daily Dow Jones Industrial Average index from 1900 to 2009 to examine the predictability of time-varying returns using automatic variance ratio-based test statistics.\footnote{Examples of test statistics include \citetapos{choi1999trw} automatic variance ratio test, \citetapos{escanciano2006gst} generalized spectral test, and \citetapos{escanciano2009apt} automatic portmanteau test.} They find compelling evidence of changes in return predictability over time, as well as an association between return predictability and both stock market volatility and economic fundamentals. This finding is consistent with the implications of the AMH. By employing the monthly S\&P composite index from 1871 to 2012 along with a generalized least squares (GLS)-based time-varying autoregressive (TV-AR) model, \citet{ito2016eme} conclude that market efficiency in the U.S. stock market changes over time, confirming the application of the AMH to the U.S. stock market.

Unlike the U.S. stock market, prior to \citet{hirayama2018erj}, there was no composite stock market index for the prewar and wartime Japanese stock market, as it differed significantly from the modern market in the following institutional aspects. First, the capital increases were mainly through a rights offer to shareholders issued at par value, while modern capital increases are mainly through public offers issued at market value (\citet{saito2016pps}). Thus, existing shareholders had to add the difference between the market value and par value  as profit or loss to their investment results. Second, in the prewar and wartime periods, capital was paid according to the part-paid stock system, which differs from the current system. Shareholders of unpaid stock were forced to make additional investments whenever additional payments were announced. This implies that additional payments had to be deducted from the investment performance in response to the capital increase. These institutional differences made it difficult to construct a composite stock market index for the prewar and wartime Japan. In other words, the contribution of \citet{hirayama2018erj} is to have constructed a composite stock market index for prewar and wartime Japan, reflecting such institutional differences. On the other hand, previous studies use various noncomposite stock market indexes to examine whether the prewar and wartime Japanese stock market was efficient in the context of the EMH.

\citet{kataoka2004b} employ daily stock prices for 1900 (from January 5 to December 29) and calculate autocorrelation coefficients, finding that the market was almost efficient in the weak sense of \citet{fama1970ecm}. \citet{suzuki2012pwt} uses the breakpoints estimated by \citet{choudhry2010ww2} and daily stock market data to investigate their relationship with major historical events during the Pacific War and the variation in stock prices.\footnote{\citet{suzuki2012pwt} focuses on the following major historical events during WWI\hspace{-.1em}I: (1) the attack on Pearl Harbor on 7 Dec 1941; (2) the Japanese conquest of Burma from January to May 1942; and (3) the Battle of Midway in June 1942.} \citet{suzuki2012pwt} concludes that market efficiency declined after the beginning of the Pacific War because the war risk premium of \citet{reitz1988erp} and \citet{barro2006rda} was reflected promptly in the price formation process. \citet{bassino2015iet} use the simple arithmetic average of daily stock prices from 1931 to 1940, along with \citetapos{engle1987etv} generalized autoregressive conditional heteroskedasticity-in-mean (GARCH-in-mean) model to investigate information efficiency in the prewar and wartime Japanese stock market. Although they find that the Japanese stock market deviated from weak-form efficiency in the 1930s, their price index does not include the shares of the Tokyo Stock Exchange (TSE), which accounts for the largest shares in terms of trading volume, and thus, does not appropriately represent the price trends of the prewar and wartime Japanese stock market. Therefore, the datasets and econometric methods used in the previous studies are both unsophisticated. Moreover, the sample periods in most studies are too short to examine whether the modern stock market has evolved and market efficiency changed over time, in accordance with the AMH. Clearly, these price indexes do not accurately reflect stock market performance.

The prevailing view in Japanese economic history is that the stock market stagnated during wartime because of wartime controls. Indeed, as \citet{okazaki1999cg} points out, based on money flow data, the corporate funding function of the stock market has regressed. However, even after the 1930s, when wartime economic controls tightened, stock market volatility and liquidity remained at a certain level. As for the wartime period, \citet{hirayama2018erj,hirayama2022smf} point out that the volatility of stock returns was still over 0.11 (from January 1940 to November 1944), while the variation in returns in the Japanese government bond market disappeared.\footnote{The average volatility in stock returns was over 0.16 from July 1924 to November 1944.} Therefore, we cannot assume that the pricing function in the stock market had disappeared. At the time, the Japanese stock market, which mostly cleared futures transactions that did not require cash equivalents as \citet{hamao2009lpd} pointed out, could keep its pricing function separate from the flow of funds . Thus, we consider the functions of corporate financing and pricing in the stock market separately and that market efficiency must be verified for the functions of pricing in the stock market during wartime. Furthermore, as \citet{suzuki2012pwt} points out, the existence of a war risk premium increased stock price volatility in wartime Japan, which in turn reduced market efficiency; thus, it is necessary to examine market efficiency given investors' risk attitudes.

In other words, this study examines the interrelationship between the time-varying nature of market efficiency and historical events in the prewar and wartime Japanese stock market from the perspective of the AMH. Our examination is fourfold. First, we provide a historical review of the datasets used in previous studies from the viewpoint of sample periods and data adequacy, and we pinpoint the problems of these datasets. Second, we introduce an accurate performance index of the prewar and wartime Japanese stock market from 1924 to 1943 constructed by \citet{hirayama2018erj}. We confirm that a new dataset reflects the price behavior of the Japanese stock market in the prewar and wartime periods accurately.\footnote{The period analyzed in this study is from June 1924 to June 1943, when deferred fee data for short-term clearing transactions are available.} Third, we examine whether the modern stock market evolved and market efficiency changed over time in accordance with \citetapos{lo2004amh} AMH. At this stage, we measure the degree of market efficiency, taking into account investors' risk attitudes, with statistical inferences using \citeapos{ito2014ism}{ito2014ism,ito2022aae} GLS-based time-varying parameter model. This approach is independent of sample size and, thus, superior to conventional statistical tests used in previous studies, such as \citetapos{kim2011srp} automatic variance ratio test using moving-window samples. Fourth, we investigate the relationships between major historical events and variations in market efficiency using one of the most reliable primary sources---the Bank of Japan's {\it{Monthly Survey}}---which is more reliable than other sources for interpreting variations in market efficiency, because it is written in a unified format and ensures data continuity. We reliably detect when the efficiency of the prewar and wartime Japanese stock market increased or declined over time, and thus, investigate the possibility that historical events have affected market efficiency.

The remainder of this paper is organized as follows. Section \ref{prewar_stock_sec2} provides a historical review of the prewar and wartime Japanese stock market. Section \ref{prewar_stock_sec3} presents the empirical methodology for estimating the degree of market efficiency based on \citeapos{ito2014ism}{ito2014ism,ito2022aae} GLS-based time-varying parameter model. Section \ref{prewar_stock_sec4} introduces the characteristics of the new dataset of price indexes for the prewar and wartime Japanese stock market constructed by \citet{hirayama2018erj}, and shows the results of some statistical tests. Section \ref{prewar_stock_sec5} presents the empirical results obtained using the GLS-based time-varying parameter model. Section \ref{prewar_stock_sec6} discusses the relationships between major historical events and time-varying market efficiency in the prewar and wartime Japanese stock market. Section \ref{prewar_stock_sec7} concludes.

\section{A Historical Review}\label{prewar_stock_sec2}
In this section, we describe the characteristics of the prewar and wartime Japanese stock market, including the types and methods of transactions and the stock market index.

\subsection{Preliminaries}
In 1878, the TSE and the Osaka Stock Exchange (OSE) were established, and there were more than 40 stock exchanges in major cities across Japan, after which they had consolidated to about 10 by c.1910. The TSE's trading share was high, generally over 40\% since 1904. The TSE is a representative stock exchange in Japan because of its large number of listed stocks, making it suitable for testing stock market efficiency. The prewar and wartime Japanese stock market differed from the present stock market in three ways: it was a part-paid stock system, there was a concentration of trading in specific stocks, and the focus was on futures trading. The part-paid stock system is described in the introduction. The remaining two unique aspects of the prewar and wartime Japanese stock market are discussed in the following subsections.

\subsection{Stock Market Centered on Nondeliverable Trading}
\begin{figure}[tbp]
 \caption{Trading Volume Share on the Market by Type of Transaction}
 \label{prewar_stock_fig1}
 \begin{center}
 \includegraphics[scale=0.7]{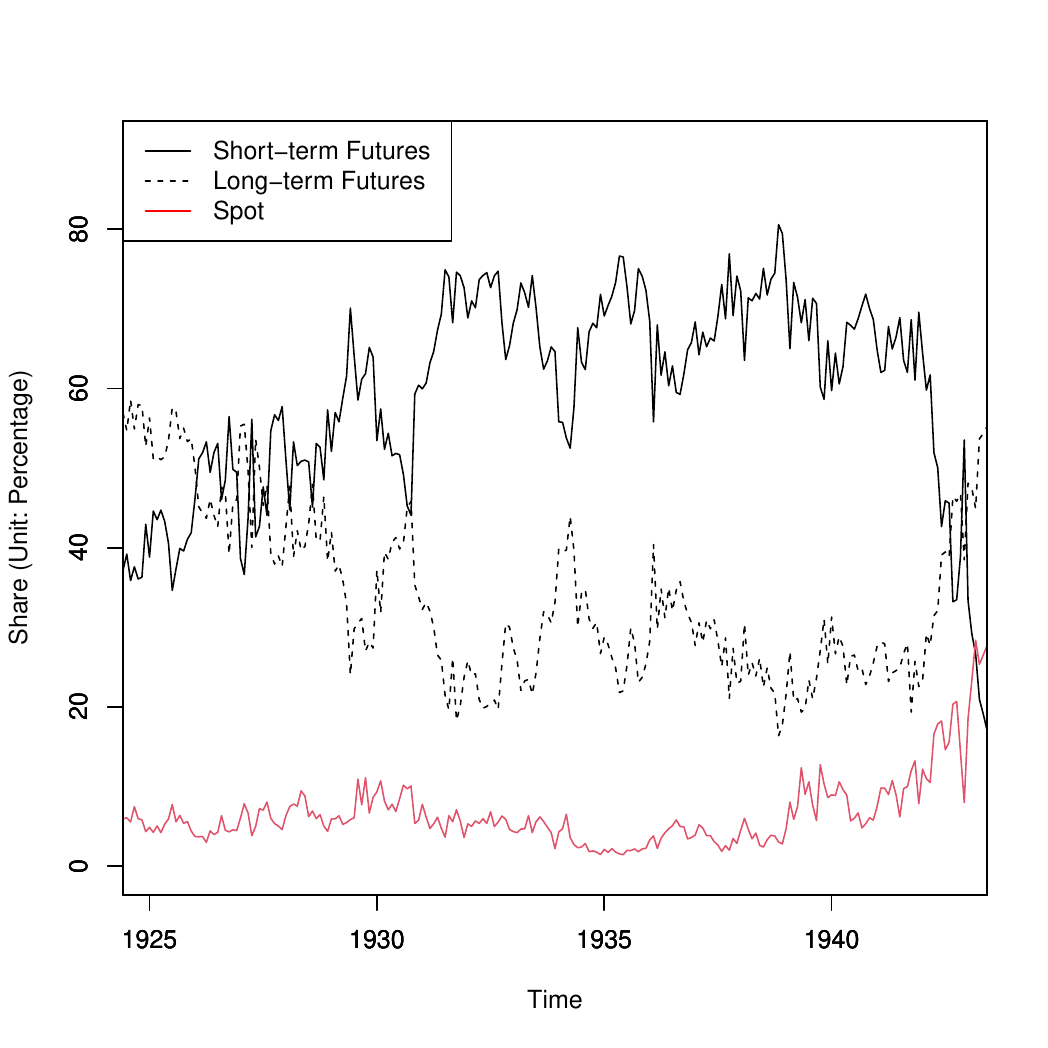}
\vspace*{3pt}
{
\begin{minipage}{400pt}\footnotesize
\underline{Source}: \citet{tse1938htse} and {\it{The Monthly Statistic Report of Tokyo Stock Exchange}}, June 1924--June 1943.
\end{minipage}}%
\end{center}
\end{figure}

In the prewar period, there were three trading methods for stock exchanges: spot transactions (jitsubutu torihiki), long-term clearing futures transactions (choki seisan torihiki; one-, two-, and three-month contracts), and short-term clearing futures transactions (tanki seisan torihiki; seven-day contracts). While spot transactions assume stock delivery, long- and short-term clearing futures transactions focus on nondeliverable trading. As Figure \ref{prewar_stock_fig1} shows, spot transactions did not exceed 10\% of all transactions on the TSE, the largest stock exchange in prewar Japan. The TSE itself was a listed stock company until the early 1940s, with futures transactions accounting for the majority of transactions. The main trading activity on these exchanges was not spot trading of individual stocks or futures trading of stock indexes, but futures trading of individual stocks. \citet{hamao2009lpd} note that, in a world context, it was rare to trade individual stocks in futures. The history of Japanese futures trading is rooted in rice trading at Dojima, Osaka, and it can be assumed that futures trading has traditionally been accepted in Japan as a commercial practice.
\begin{center}
(Figure \ref{prewar_stock_fig1} around here)
\end{center}

In particular, the proportion of transactions related to the TSE's new shares was high, such as major issues of short-term futures; furthermore, the share of short-term futures transactions rose to 81\% in November 1938. Therefore, the market was dominated by highly liquid short-term futures transactions until the end of the 1930s. In this study, we calculate the stock index for the TSE as an exchange; therefore, we should be careful not to confuse it with TSE's new shares as individual stocks. The positioning of TSE's new shares in long- and short-term clearing trades on the TSE is as follows. From 1928 to 1942, the average market capitalization ratio was only 1.1\%, but trading volume accounted for 45.5\%. Therefore, TSE's new shares should be considered when studying prewar and wartime stock markets. At the same time, if we place too much emphasis on the trading volume share, it becomes difficult to observe the characteristics of other stocks, and we should proceed with caution. In December 1941, when the Pacific War began, the share of short-term futures transactions declined.\footnote{Finally, the TSE was unlisted on 31 August 1943, and the exchanges in various regions were consolidated into the Japan Stock Exchange. Accordingly, short-term clearing futures transactions were abolished, and stock transactions were conducted through long-term clearing futures and spot transactions.}

While shares were delivered for spot transactions, the delivery rate was low for long- and short-term clearing futures transactions because most were net settlements. An increase in nondeliverable trading relative to settlements associated with the transfer of shareholders can be considered an indicator of an increased market pricing function and liquidity. Nondeliverable trading provides market participants with a means of buying and selling without financial constraints, thereby contributing to flexible pricing and smooth trading.

\subsection{Methods for Stock Price Index Calculation and Market Characteristics in the Early Showa Period}\label{prewar_stock_sec2-2}
Short-term clearing futures transactions constituted the core of the Japanese stock market during the early Showa Period.\footnote{We define the period of prewar and wartime Japan as the ``early Showa period.''} The initially listed stocks were limited to TSE's and Kanegafuchi Spinning's new shares in June 1924. As other issues were added to short-term clearing futures transactions, the effect of the trading of two specified stocks on the market capitalization-weighted average index decreased. However, as the trading volume of these stocks accounted for the majority of transactions until 1942, the effect of specific share trading on the index weighted by trading volume remained strong. A stock index that covers many stocks can be classified into three types of weighting methods: equal-weighted (arithmetic) average index, trading-volume weighted average index, and market capitalization-weighted average index. There is a difference in the characteristics of stock markets based on a market capitalization index versus a trading volume index. Notably, most stock indexes created during the early Showa period were trading volume-weighted averages or simple arithmetic averages, whereas the modern stock index is generally based on the market capitalization-weighted average.

The trading volume-weighted average was significantly affected by the price fluctuations of some stocks with high trading volumes (the TSE's and Kanegafuchi Spinning's new shares) to represent the overall performance of the stock market. For example, the {\it{Kabuka Dai-Shisu}} (stock price index) calculated by the TSE using \citetapos{fisher1921bfi} formula, which is one of the methods for calculating the consumer price index according to transaction volume, and the {\it{Zen-Sango Shin-Kyu Sogo Kabuka Shisu}} (all-industry old and new composite stock price index) conducted by \citet{fujino1977sir} are weighted average indexes based on trading or settlement volume. Therefore, the effect of TSE's new shares was significant, and the effect of excessively concentrated trading could not be eliminated. Consequently, to understand the characteristics of the overall stock market, it is necessary to dilute the excessive effects of some stocks such as TSE's new shares.

Meanwhile, the equally weighted average index reduces the impact of stocks with extremely high trading volumes but has the disadvantage of reflecting excessively small and illiquid stocks in the index. The stock price index calculated by the Bank of Japan and Daily Stock and Government Bond Price Index calculated by {\it{Toyo Keizai Shimpo}}, adopted by \citet{bassino2015iet}, are based on simple arithmetic averages. Thus, they can be used to eliminate the effects of TSE's new shares. However, because the ratio of small-cap stocks was relatively high, the impact of large-cap stocks may have been underestimated.

Spot and long-term clearing futures transactions involve many traded stocks with high market coverage. However, the trading volume per stock was low and the number of stocks with consecutive months of no trading was too high to ignore. In the case of short-term futures transactions, the volume of trading per stock was large, and there were few consecutive months without trading. This led us to recognize that short-term clearing futures transactions had a more effective function in stock price determination than spot and long-term clearing futures transactions. Therefore, this study examines market efficiency using the equity performance index (EQPI), a market capitalization-weighted index for short-term clearing futures transactions. There was a unique capital system in the prewar and wartime Japanese stock market in which purchasers of company stock issues paid the par value of their shares in installments rather than in full, called the ``Part-Paid Stock System.'' The EQPI are the only indexes adjusted for ex-rights, additional payments, and dividends based on this system.

As described above, the prewar stock index in Japan reflects a market structure in which the market characteristics are easily influenced by the calculation method used. Since the mid-1930s, the Japanese government has strengthened its wartime production system, and the Japanese economy has become susceptible to the performance of the munitions industry. Simultaneously, several manufacturing companies belonging to the munitions industry were added to the list of short-term clearing futures transactions. Consequently, the market capitalization ratio of such sectors has also increased, and the characteristics of the EQPI are likely to be more influenced by domestic policy and corporate performance in manufacturing industries. Subsequently, the Bank of Japan's {\it{Monthly Survey}} showed that stock market participants paid attention to these changes after the mid-1930s.

First, to confirm whether the stock market's characteristics have changed, it is important to identify the market's characteristics. The {\it{Monthly Survey}}'s comments on market trends provided insight into the changing perspectives of market participants. The {\it{Monthly Survey}} is a more reliable data source than newspapers because it is a market survey conducted by a public organization, is presented in a consistent format, and has been conducted over a long period. The sample period is 203 months, from June 1924 to April 1941. We classify the stock market comments in the {\it{Monthly Survey}} into 14 market factors, and the secular changes are shown in Table \ref{prewar_stock_table1}.\footnote{Market trends have not been reported in the {\it{Monthly Survey}} since May 1941.}
\begin{center}
(Table \ref{prewar_stock_table1} around here)
\end{center}

Table \ref{prewar_stock_table1} shows the changes in the factors around the stock market on which market participants focus when making trading decisions. Until 1933, the focus was on the following factors: domestic financial market, domestic commodity market, foreign exchange, and overseas political events. For example, during the Showa financial crisis of 1927 and the monetary easing between 1932 and 1933, the domestic financial market attracted significant attention as a cause of stock market fluctuations. The appreciation of the Japanese Yen after the lifting of the gold ban in 1930 exerted downward pressure on domestic commodity prices, and the domestic commodity market factor influenced stock prices. However, the depreciation of the Japanese Yen following the rebanning of gold exports in 1932 influenced stock prices through export performance, thereby increasing the focus of market participants on the foreign exchange factor. Until the early 1930s, for short-term clearing futures transactions, there was a high ratio of exchange stocks sensitive to monetary policy to spinning stocks sensitive to foreign exchange and commodity prices. Here, the EQPI is a market capitalization-weighted index, which well represents the composition ratio of traded stocks and, thus, can adequately reflect the fluctuating factors perceived by market participants.

Meanwhile, in the process of strengthening wartime controls after the mid-1930s, market participants' attention shifted to the following factors: domestic policy, overseas political events, and domestic corporate performance. From 1934 to 1936, the corporate profits of manufacturing industries, such as machinery, metals, and chemicals, recovered, and stock prices were increasingly based on corporate performance. In addition, after 1936, domestic policy and overseas political events received increasing attention, and policies on economic control and wartime mobilization influenced the stock market. The composition of the EQPI was likely to be more reflective of the interest of market participants, as manufacturing industrial stock prices rose markedly after the mid-1930s and the composition ratio of securities exchange stocks fell sharply after the mid-1930s.

Next, we highlight the appropriateness of the EQPI calculated for short-term clearing futures transactions. The margins (deferment fees) for trading in these futures transactions fluctuate daily, and stock prices are determined not only by supply and demand in the stock market, but also by the external political and economic environment. Deferment fees can be considered as a margin for individual stock futures that change daily. The extreme changes in deferment fees, as frequently noted in the {\it Monthly Survey}, influenced market participants' trading motives and stock prices. We use the EQPI constructed by \citet{hirayama2018erj}, because an appropriate stock index reflecting market fluctuations should be used to examine market efficiency. In the short-term clearing futures transactions adopted by the EQPI, the timing of changes in the supply and demand of the stock market, indicated by the rise and fall in deferment fees, coincides with the timing of the major factors. In other words, we should consider the possibility that stock prices and deferment fees were determined simultaneously when examining market efficiency in the prewar and wartime Japanese stock market. In the following section, we explain this point in terms of deferment fees and confirm their specific relationship with stock prices.

\begin{figure}[tbp]
 \caption{Deferment Fee on Short-Term Futures Transactions (All Stocks Weighted)}
 \label{prewar_stock_fig2}
 \begin{center}
 \includegraphics[scale=0.7]{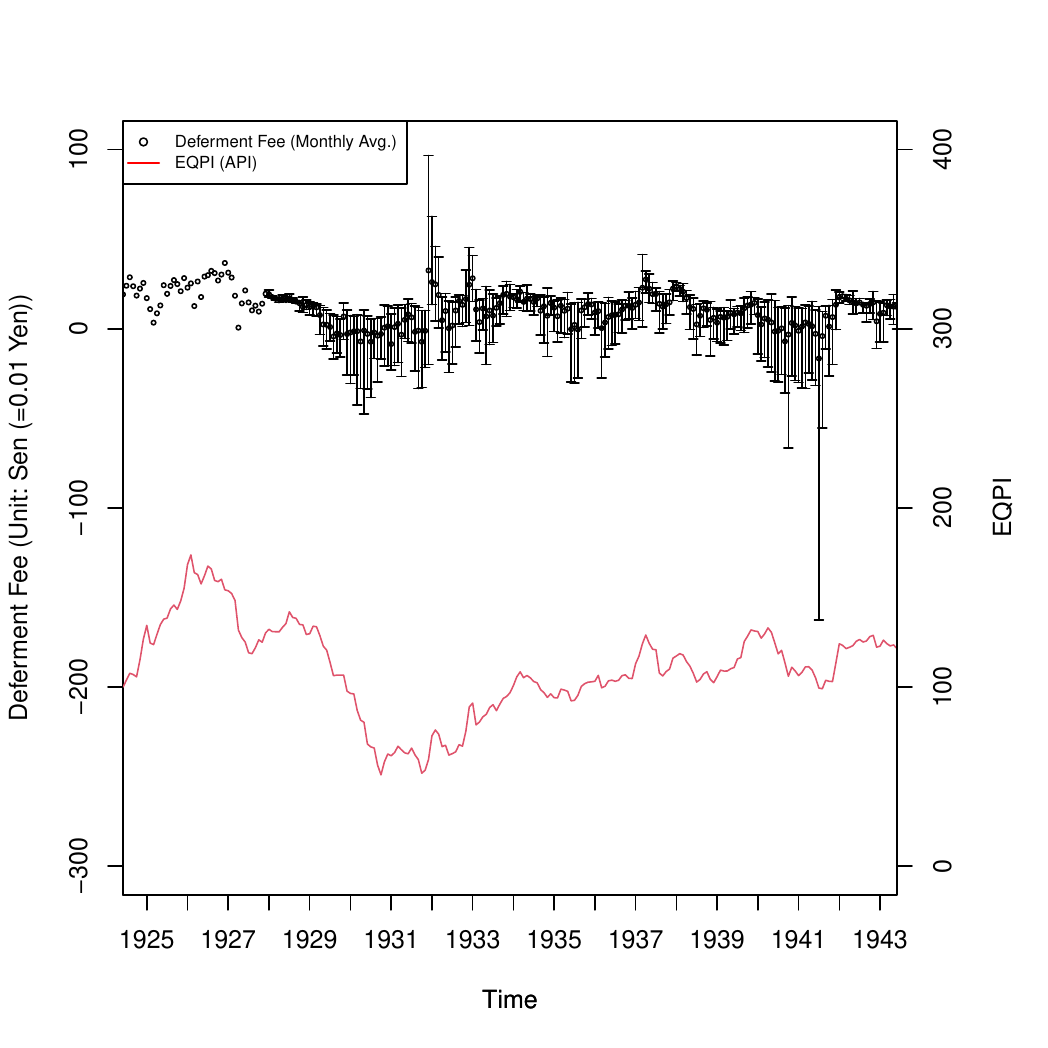}
\vspace*{3pt}
{
\begin{minipage}{400pt}\footnotesize
\underline{Source}: As for Figure \ref{prewar_stock_fig1}. 
\end{minipage}}%
\end{center}
\end{figure}

A rule for short-term clearing futures transactions was to transfer payments and share certificates simultaneously with trading (same-day delivery for transactions processed on the previous day's afternoon and the current day's morning); however, it was possible to defer delivery and settlement through a deferment fee. These transactions were essentially the same as the indefinite difference settlement transactions. The role of the ``deferral agencies'' in facilitating them should be noted. The deferral agencies executed delivery on behalf of all sellers and buyers. Specifically, when excessive shares had to be delivered, payment was made on behalf of the purchaser, and excess shares were received (cash transfer); when there was a shortage of shares, the shortfall was covered on behalf of the seller, and payment was received by the agency (stock transfer). Because the deferment agency decided on the deferment fee based on the distortion of the trading position, this level was regarded as one of the indicators of the overall market situation. Specifically, the deferment fee was determined by the market interest rate, deferral conditions of the trading member, and the agency's willingness to adjust delivery. The decisive factor was the amount of stock that could be delivered. Deferment agencies intentionally increased deferment fees to adjust for bias in deliveries deferred by market participants. For this reason, it was assumed that the positive deferment fee would be set at a high rate when the number of shares on the buy side increased rapidly, and the negative deferment fee would be set at a high rate when the number of shares on the sell side increased rapidly, as \citet{hirayama2022smf} pointed out.
\begin{center}
(Figure \ref{prewar_stock_fig2} around here)
\end{center}
\noindent Figure \ref{prewar_stock_fig2} shows the weighted average of the deferment fees for all stocks in the short-term clearing futures transactions market. The average monthly deferment fee for the period examined in this study is approximately 11 sen ($=0.11$ yen), lower than the TSE's new shares (20 sen). In the early Showa period, negative deferment fees were observed for all short-term clearing futures transactions (weighted average market capitalization), mainly in two periods: (1) from the Showa Depression to the lifting of the ban on gold exports (from August 1929 to October 1930); and (2) around July 1941, when the U.S. President froze all Japanese assets in the U.S.

Meanwhile, positive deferment fees rose sharply in the following four periods: (1) in December 1926, when buying speculation with respect to buying Kuhara Mining shares reached its peak after November 1926; (2) after December 1931, when the new Cabinet of Prime Minister Tsuyoshi Inukai (Minister of Finance Korekiyo Takahashi) promulgated the ban on gold exports and announced additional payment of the TSE's new shares, causing the stock market to soar (from December 1931 to March 1932); (3) from December 1932 to January 1933, when expectations of monetary easing grew and stock exchanges recorded the trading volume after the Bank of Japan began underwriting government bonds amid the appreciation of the U.S. dollar; and (4) after March 1937, when the TSE as a stock exchange recorded its highest trading volume before the war. When the positive deferment fee was high, the fluctuations were limited. As an exception, in December 1931, it reached 97 sen ($=0.97$ yen). Thus, it can be assumed that a temporary increase in the deferment fee was implemented to tackle panic buying. Therefore, the information above indicates that in the early Showa period, a concentration of unidirectional transactions occurred around the Showa depression during a series of policy changes to lift the ban on and reprohibit gold exports.

As described above, in addition to the characteristics of the stock index, the market characteristics of the prewar period were confirmed multilaterally from such aspects as the main sector transitions based on monthly surveys and the perceptions of market participants, as well as changes in supply and demand distortions based on deferred changes for short-term clearing futures transactions. The results show that it is appropriate to examine market efficiency using a market capitalization-weighted average stock index for short-term clearing futures transactions. Next, we examine whether market conditions impact stock market efficiency.

\section{The Model}\label{prewar_stock_sec3}
This section reviews our empirical method for examining the efficiency of the prewar and wartime Japanese stock market, which is considered an established modern financial market. Our method is based on \citeapos{ito2014ism}{ito2014ism,ito2022aae} GLS-based time-varying parameter model. Unlike the conventional model used in previous studies, this approach aims to investigate market efficiency by considering time-varying interrelationships among variables, such as the relationship between stock returns and investors' risk attitudes. In practice, we use a special case of their model, a GLS-based time-varying vector autoregressive (TV-VAR) model, to estimate the time-varying degree of market efficiency in the prewar and wartime Japanese stock market for each period using multivariate data.

\citet{fama1970ecm} argues that security prices immediately reflect instantaneous shocks if the market is efficient, implying that such shocks do not propagate persistently to stock returns and investors' risk attitudes. Thus, our approach is based on impulse responses using multivariate time-series data. The impulse response reflects the propagation of an instantaneous shock to the system and the aggregate impulse response represents the full effect of the shock. To this end, we first employ a conventional vector autoregressive (VAR($p$))  model to obtain impulse responses for multivariate data and analyze the interrelationships among the variables in the context of market efficiency. Let ${\ve{y}}_t$ denote a vector representing stock returns and changes in the deferment fee at $t$:
\begin{equation}
 \ve{y}_t=\ve{\nu}+A_1\ve{y}_{t-1}+\cdots+A_p\ve{y}_{t-p}+\ve{u}_t; \ \ t=1,2,\ldots,T, \label{eq1}
\end{equation}
where $\ve{\nu}$ is a vector of constant terms and $\ve{u}_t$ is a vector of error terms, following an independent and identically distributed multivariate process with a mean of zero vectors. When ${\ve{y}}_t$ is stationary, we can invert Equation (\ref{eq1}) into the following vector moving average (VMA ($\infty$)) model using Wald decomposition:
\begin{eqnarray}
\ve{y}_{t}&=&\ve\mu+\Phi_{0}\ve{u}_{t}+\Phi_{1}\ve{u}_{t-1}+\Phi_{2}\ve{u}_{t-2}+\cdots\nonumber\\
&=&\ve\mu+\Phi(L)\ve{u}_t, \label{eq2}
\end{eqnarray}
where $\Phi(L)$ is a matrix lag polynomial of lag operator $L$ with all eigenvalues of $\Phi(1)$ outside the unit circle, and $\ve\mu$ is the mean of $\ve{y}_t$. We assume that the coefficient matrixes $\{\Phi_i\}^\infty_{i=0}$ are $k\times k$-dimensional ($2\times 2$-dimensional, in our case) parameter matrixes. We then compute the impulse response functions along with identification assumptions, such as $\Phi_0=I$, an identity matrix. This means that the long-run effect of $\ve{u}_t$ on $\ve{y}$ is given by $\Phi(1)$; the vector of expected returns on stock and deferment fee, $\ex\left[{\ve{y}}_t \ | \ {\ve{y}}_{t-1},{\ve{y}}_{t-2},\cdots\right]-\ve{\mu}$, is zero when $\Phi(1)=I$. In other words, the stock market is efficient when $\Phi(1)=I$ or $\left[I-A(1)\right]^{-1}=\left[I-A_1-A_2-\cdots-A_p\right]^{-1}=I$ is satisfied, because there are no excess returns on the stock and no deferment fee.

Here, we construct the degree of market efficiency based on the impulse response to investigate whether the stock market was efficient in prewar and wartime Japan. The impulse response can be easily obtained by using a conventional VAR($p$) model and algebraically computing its estimates. We measure the relative degree of market efficiency that varies over time, according to the following procedure. First, we compute the cumulative sum of the coefficient matrixes of the impulse response:
\begin{equation}
\Phi(1)=\left(I - A_1 - A_2 - \cdots - A_p\right)^{-1},\label{eq3}
\end{equation}
Then, to measure the deviation from the efficient market, we define the degree of market efficiency as
\begin{equation}
\zeta=\sqrt{\mbox{max}\ \lambda\left[(\Phi(1) - I)'(\Phi(1) - I)\right]},\label{eq4}
\end{equation}
where ``max $\lambda[X]$'' denotes the maximum eigenvalue of a matrix $X$. In the case of an efficient market, where $A_1=A_2=\cdots=A_p=0$, the degree $\zeta$ becomes zero; otherwise, $\zeta$ deviates from zero. Hence, we regard $\zeta$ as the degree of market (in)efficiency. We consider a large deviation of $\zeta$ from $0$ as evidence of market inefficiency.

Next, we estimate the GLS-based TV-VAR coefficients for each period to obtain the corresponding degree defined in Equation (\ref{eq4}). Following this method, we use a model in which all VAR coefficients, except that corresponding to the constant terms $\ve{\nu}$, follow multivariate random walk processes. Thus, we assume the following:
\begin{equation}
A_{i,t}=A _{i,t-1}+V_{i,t}, \ \ i=1,2,\cdots,p, \ {\rm{and}} \ t=1,2,\cdots,T, \label{eq5}
\end{equation}
where $\{V_{i,t}\}$ is a $p\times t$-dimensional error term matrix. We assume that the matrix satisfies $E\left[V_{i,t}\right]=\ve{O}$ for all $t$, \ $E\left[vec(V_{i,t})'vec(V_{i,t})\right]=\sigma_v^{2} I$ and $E\left[vec(V_{i,t})'vec(V_{i,t-m})\right]=\ve{O}$ for all $i$ and $m\neq 0$. \citeapos{ito2014ism}{ito2014ism,ito2022aae} method allows us to estimate the following GLS-based TV-VAR model:
\begin{equation}
\ve{y}_{t}=\ve\nu+A_{1,t}\ve{y}_{t-1}+A_{2,t}\ve{y}_{t-2}+\cdots+A_{p,t}\ve{y}_{t-p}+\ve{u}_{t}, \label{eq6}
\end{equation}
along with Equation (\ref{eq5}). In this specification, we can invert Equation (\ref{eq6}) into the following time-varying VMA model:
\begin{equation}
\ve{y}_{t}=\ve\mu_t+\Phi_{0,t}\ve{u}_{t}+\Phi_{1,t}\ve{u}_{t-1}+\Phi_{2,t}\ve{u}_{t-2}+\cdots, \label{eq7}
\end{equation}
where $\Phi_{0,t}=I$ for all $t$. In addition, we modify the condition that the vector of expected excess returns on stock and deferment fee, $\ex\left[{\ve{y}}_t \ | \ {\ve{y}}_{t-1},{\ve{y}}_{t-2},\cdots\right]-\ve{\mu}$, is zero in Equation (\ref{eq2}) as $\ex\left[{\ve{y}}_t \ | \ {\ve{y}}_{t-1},{\ve{y}}_{t-2},\cdots\right]-\ve{\mu}_t$ where $\ve{\mu}_t=(I-A_{1,t}-\cdots-A_{p,t})^{-1}{\ve{u}}_t$. We can then derive the time-varying degree of market efficiency, because the TV-VAR model can be considered a locally stationary model. In practice, we first compute the cumulative sum of the time-varying VMA coefficient matrixes that appear in Equation (\ref{eq7}).
\begin{equation}
 \Phi_t(1)=\sum_{j=1}^{\infty}\Phi_{j,t}, \label{eq8}
\end{equation}
As mentioned above, we find that the stock market is efficient when $\Phi_t(1)=I$ each time $t$. We do not need any identification assumptions about the long-run effect of shocks $\left\{\ve{u}_{\tau}\right\}_{\tau=t}^{\infty}$ on $\left\{\ve{y}_t\right\}$. We then define the time-varying degree of market efficiency as the distance between $\Phi_t(1)$ and $I$ measured by the spectral norm:
\begin{equation}
\zeta_t=\sqrt{\mbox{max}\ \lambda\left[(\Phi_t(1)-I)'(\Phi_t(1)-I)\right]}.\label{eq9}
\end{equation}
We call $\zeta_t$ the time-varying degree of market efficiency, which  measures how close or far from efficient markets the actual markets are at $t$ (the stock market is efficient if $\zeta_t$ is zero).

Furthermore, we apply a residual bootstrap technique to the TV-VAR model to make a statistical inference on the time-varying degree of market efficiency. Based on the hypothesis that all the TV-VAR coefficients are $0$, we formulate a set of bootstrap samples for the TV-VAR estimates. Assuming that the processes for all variables are generated in an efficient market, this procedure provides a (simulated) distribution of the estimated TV-VAR coefficients. We then compute the corresponding distributions of the impulse response and degree of market efficiency. Finally, we conduct a statistical inference of the estimates to examine whether market efficiency has changed over time in the prewar and wartime Japanese stock market. We detect periods when the prewar and wartime Japanese stock market experienced market inefficiency using confidence bands derived from such simulated distributions.

\section{Data}\label{prewar_stock_sec4}
We use the EQPI---the first market capitalization-weighted index---of the prewar and wartime Japanese stock market, calculated by \citet{hirayama2018erj}. \citet{bassino2015iet}, a representative study, use the simple arithmetic average of daily stock prices from the Japanese yearbooks of {\it{Toyo Keizai Shimpo}} ({\it Oriental Economist}).\footnote{\citet{kataoka2004b} use the daily average stock prices of the spinning and railway industries obtained from the {\it Asahi Shimbun Newspaper}  and \citet{suzuki2012pwt} calculates and employs the daily volume-weighted average index from {\it Chugai Commercial Newspaper}.} As mentioned in Section \ref{prewar_stock_sec2-2}, a simple arithmetic average of stock prices is not suitable for capturing the entire stock market adequately. For the same period as \citet{bassino2015iet}, from January 1930 to December 1940, the return on the EQPI was 0.056 for the price index (PI) and 0.057 for the adjusted price index (API). This difference is small per year (0.001), which means that the impact of the correction of ex-rights and additional payments was moderate. On the other hand, the return on the EQPI was 0.021 for the PI and 0.044 for the API in the subsequent period from January 1941 to June 1943, we could not ignore the difference (0.023). Thus, the performance of Japanese equity investments in the prewar and wartime periods required various adjustments that had different effects on prices at different times, a point that should not be taken lightly when examining market efficiency.

The EQPI is a market capitalization-weighted average that adjusts the share price by issuing new shares or making other payments.\footnote{See online Appendices A.1 and A.2 (\url{https://at-noda.com/appendix/prewar_stock_appendix.pdf}) for the significance of the EQPI and the features of short-term clearing futures transactions, respectively.} \citet{hirayama2018erj} calculates not only the PI and API, but also the total return index (TRI), which considers the effect of dividends. Ideally, the TRI is desirable for expressing investors' stock investment performance.
\begin{center}
(Table \ref{prewar_stock_table2} around here)
\end{center}
Table \ref{prewar_stock_table2} shows the differences between the well-known stock price indexes and the EQPI. The TSE uses volume-weighted averages to calculate its stock index. In addition, the Toyo Keizai Shimpo and Nippon Kangyo Bank used simple averages to calculate their stock indexes, but there is no capitalization-weighted average index, the same method used today. Furthermore,  no stock index considers the impact of new share allocations or other corporate actions (including dividends) on investment performance. We can understand the value of capital in the prewar and wartime Japanese stock market using the capitalization-weighted index. We believe that the EQPI is appropriate for studying the prewar and wartime Japanese stock market based on the following four points: (1) the TSE's new shares, which represent the prewar and wartime Japanese stock market, are subject to the EQPI; (2) the EQPI is a capitalization-weighted index, which is also used in modern stock indexes, making it more comparable to modern ones; (3) the EQPI is an index in which the impact of new share allocations and additional payments on investment performance has been appropriately corrected; and (4) the EQPI is a TRI, which includes dividends and accurately reflects investment performance.

We use three types of monthly average price indexes for the EQPI: the PI, API, and TRI. In addition, we employ deferment fees obtained from the {\it{Monthly Statistics Report of the Tokyo Stock Exchange}} as a proxy variable for investors' risk attitudes, because if the market is efficient, then deferment fees would be efficiently determined. The sample period for each variable is the same, from June 1924 to June 1943. We take the log differences in the price indexes and fees to obtain stock returns and changes in fees.
\begin{center}
(Table \ref{prewar_stock_table3} around here)
\end{center}
Table \ref{prewar_stock_table3} presents the descriptive statistics of the variables. The mean of the returns on the TRI is higher than that on the PI and API because of adding income gains. The total returns were heavily dependent on income gains in the prewar and wartime Japanese stock market, thus we should take dividends into account. Meanwhile, the volatility of the changes in deferment fees is extremely high, which means that the deferment agency reacted sensitively to changes in the market environment and often changed its risk attitude. We believe that this can be used as a proxy variable for investors' risk sensitivity, as changes in deferment fees tend to expand when supply and demand in the short-term clearing market become more distorted. Thus, we should allow for the possibility that changes in deferment fees are determined simultaneously with each stock return in the VAR system.

Table \ref{prewar_stock_table3} presents the results of the unit-root test using descriptive statistics. For the estimation, all the variables that appear in the moment conditions should be stationary. We apply the augmented Dickey--Fuller GLS (ADF-GLS) test of \citet{elliott1996eta} to confirm whether the variables satisfy the stationarity condition. Furthermore, we employ a modified Bayesian information criterion instead of the modified Akaike information criterion to determine the optimal lag length. This is because the estimated coefficient of the detrended series $\hat\psi$ does not indicate the possibility of size distortions \cite[]{elliott1996eta,ng2001lls}. The ADF-GLS test rejects at the 1\% significance level the null hypothesis that all the variables contain a unit root.

\section{Empirical Results}\label{prewar_stock_sec5}

\subsection{Preliminary Estimations}
First, we assume a time-invariant VAR($q$) model with a constant and employ \citetapos{schwarz1978edm} Bayesian information criteria to determine the optimal lag order for the preliminary estimations. Table \ref{prewar_stock_table4} summarizes the results of the time-invariant VAR($q$) model. We employ bivariate second-order vector autoregressive (VAR($2$)) models for our time-invariant estimations.
\begin{center}
(Table \ref{prewar_stock_table4} around here)
\end{center}
Table \ref{prewar_stock_table4} shows that all estimates, except those for the changes in the deferment fee in the equation for each stock return, are significant at the 1\% level. We find that the absolute values of the cumulative sum of the estimates of the autoregressive terms for each variable in the equation for TRI are the smallest, followed by those for API and PI. The TRI is the most efficient, because the returns follow the random walk process when the estimates are close to zero. In addition, this finding suggests that investors make decisions based not only on stock prices, but also on dividends. Based on this result and the limited explanatory power of the conventional VAR model, it is important to consider the time-varying nature of the prewar and wartime Japanese stock market efficiency.

Then, we examine whether the parameters are constant in conventional VAR($p$) models using \citetapos{hansen1992a} parameter constancy test under the random parameter hypothesis. As shown in Table \ref{prewar_stock_table4}, we reject the null hypothesis of constant parameters against parameter variation as a random walk at the 1\% significance level. The results suggest that the conventional VAR($p$) model does not apply to our data, but that the TV-VAR($p$) model is a better fit. From a historical perspective, the prewar and wartime Japanese stock market experienced various exogenous shocks, such as bubbles, economic or political crises, natural disasters, policy changes, and wars. Table \ref{prewar_stock_table5} summarizes the major historical events that occurred in the prewar and wartime Japanese stock market.
\begin{center}
(Table \ref{prewar_stock_table5} around here)
\end{center}
We consider that these events affected stock price formation. In the following subsection, we estimate the time-varying degree of market efficiency using a GLS-based TV-VAR model.

\subsection{Time-Varying Degree of Market Efficiency}
\begin{figure}[hbp]
 \caption{\small{Time-Varying Degree of Market Efficiency (EQPI)}}
 \label{prewar_stock_fig3}
 \begin{center}
 \includegraphics[scale=0.4]{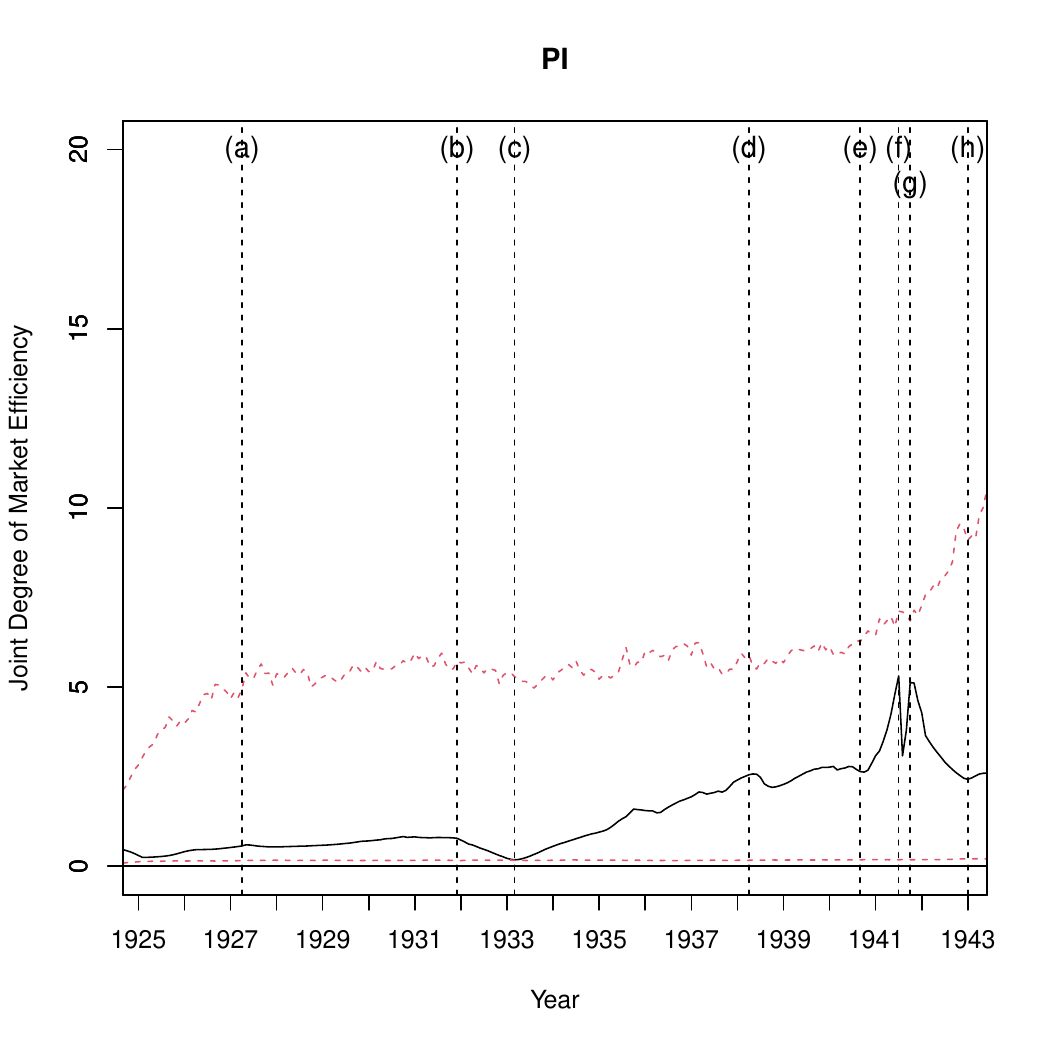}
 \includegraphics[scale=0.4]{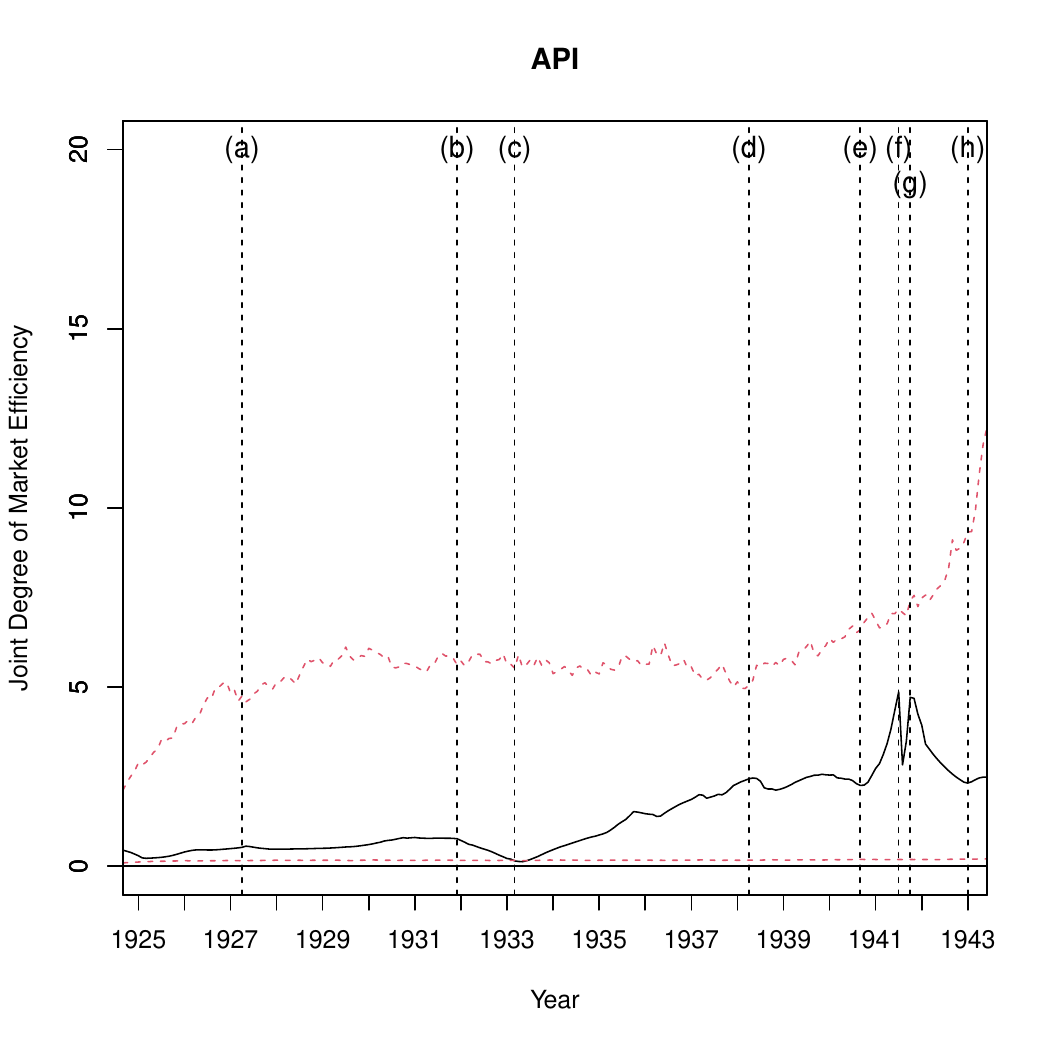}
 \includegraphics[scale=0.4]{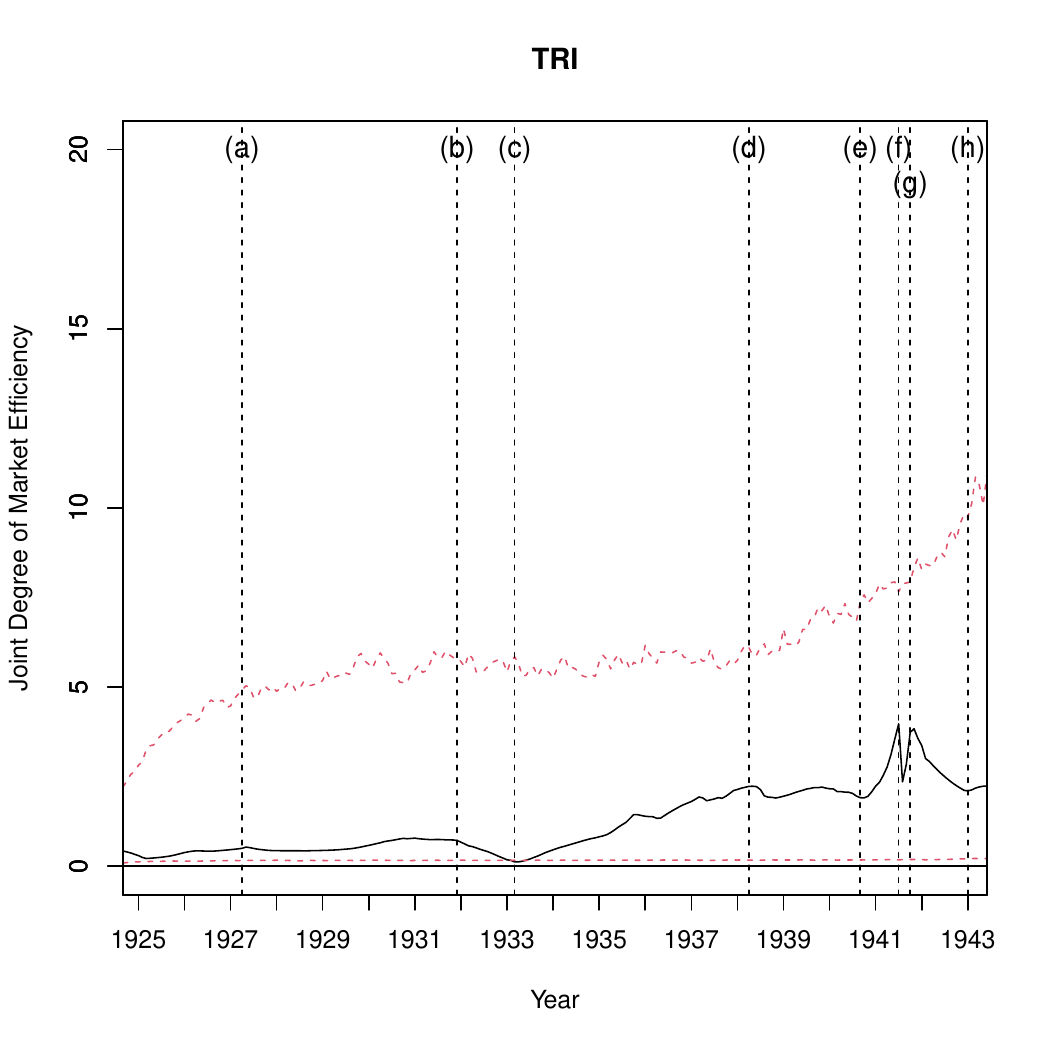}
\vspace*{3pt}
{
\begin{minipage}{400pt}\footnotesize
\underline{Notes}:
\begin{itemize}
 \item[(1)] The points (a)--(h) in the figure are consistent with the time periods in Table \ref{prewar_stock_table6}. 
 \item[(2)] The dashed red lines represent 99\% confidence intervals of the efficient market degrees.
 \item[(3)] We run the bootstrap sampling 10,000 times to calculate the confidence intervals.
 \item[(4)] R version 4.4.0 is used to compute the estimates.
\end{itemize}
\end{minipage}}%
\end{center}
\end{figure}

In this subsection, we first review market efficiency following \citet{fama1970ecm}. There are no arbitrage opportunities, because asset prices reflect all available information, including historical prices, if the market is efficient in the weak sense. However, there is no consensus on whether the prewar and wartime Japanese stock market was efficient in the weak sense \citep[]{kataoka2004b,suzuki2012pwt,bassino2015iet}.
\begin{center}
(Figure \ref{prewar_stock_fig3} around here)
\end{center}
Next, we focus on how the degree of market efficiency fluctuated over time in the prewar and wartime Japanese stock market. Again, we measure the stock market deviation from the efficient condition using Equation (\ref{eq9}). The dashed red lines in Figure \ref{prewar_stock_fig3} represent the 99\% confidence bands of the degree for an efficient market.\footnote{Thus, we construct the confidence bands for testing the null hypothesis that $\zeta_t$ is zero for each period.} In other words, the degree is not significantly different from zero if it is found within these bands, which means that the efficient condition was satisfied for the prewar and wartime Japanese stock market. Thus, the market was almost efficient in the absolute sense, but the relative market efficiency was quite volatile throughout the sample period.

In particular, market efficiency continued to deteriorate after Japan's withdrawal from the League of Nations in March 1933, and market (in)efficiency peaked in July 1941, when U.S. President Franklin D. Roosevelt declared a freeze on all Japanese assets in the U.S. Although market efficiency improved temporarily with the progress of U.S.--Japan peace negotiations, it declined again with the resignation of Fumimaro Konoe's cabinet in October 1941. Subsequently, market efficiency improved, reflecting Japanese victories in the early stages of the Pacific War. However, after the withdrawal of Japanese forces from Guadalcanal in February 1943, market efficiency began to deteriorate again. These results are consistent with earlier studies \citep[e.g.,][]{berkman2011tvr,caldara2022mgr,verdickt2020ewr}, which argue that investors asked for a ``war risk premium.''

In the following section, we focus on March 1933, when the degree of market efficiency began to change rapidly. We divide the discussion into two periods: June 1924--March 1933 and April 1933--June 1943. Based on the characteristics of the prewar and wartime Japanese stock market, such as trading volume and deferment fees, as described in Section \ref{prewar_stock_sec2}, we elucidate, from a historical perspective, the factors causing fluctuations in the degree of market efficiency.

\section{Historical Interpretation}\label{prewar_stock_sec6}
In this section, we discuss the correspondence between the variation in the degree of market efficiency and historical turning points from the perspective of the AMH. Again, the AMH asserts how market efficiency changes over time in a relative sense and does not investigate whether the market is efficient in an absolute sense, as examined in previous studies.

\subsection{The Showa Financial Crisis and Suspension of the Gold Standard in Japan (June 1924--March 1933)}
 Table \ref{prewar_stock_table6} lists the main turning points in the prewar and wartime Japanese stock market.
\begin{center}
(Table \ref{prewar_stock_table6} around here)
\end{center}
Market (in)efficiency increased during the following turning points: (a) the financial moratorium from April to May 1927, and turning point (b) the suspension of the gold standard by the Inukai Cabinet, which achieved a double-bottom pattern of market efficiency in December 1931. During these periods, market participants leaned toward a selling position in the mid-1930s from a buying position in 1932.

First, we review the degree of market efficiency during the financial moratorium period of early 1927. The TSE's short-term clearing futures transactions began in June 1924, a few months after the Great Kanto Earthquake in September 1923, and substantially affected the Japanese economy before the war. During this period, the Bank of Japan established a large-scale monetary easing policy for post-earthquake reconstruction, which positively affected the stock market. In April 1925 and October 1926, the bank reduced the discount rate for commercial bills, and the degree of market efficiency remained high during the recovery of stock prices. Unlike in the period up to the beginning of the 1920s, the frequency of new share allocations and additional payments decreased, as did the return gap between the PI and API. As Figure \ref{prewar_stock_fig3} shows, both indicators have a similar trend. From February to March 1926, stock prices increased sharply, despite the depreciation of the U.S. dollar. Stock market participants tended to value the strong yen when considering the recovery of Japan's national strength. Furthermore, stock dividend yields were revalued because of the decline in policy interest rates.

The deferment fee for short-term clearing futures transactions increased until December 1926 owing to the purchase of Kuhara Mining Co., Ltd., which accounted for 10.2\% of the market capitalization of short-term clearing futures transactions. However, the speculative stock purchases had only a partial effect on the overall market, and the degree of market efficiency based on the EQPI remained high. Next, during the financial moratorium (April--May 1927), the stock market was in a sell-off state, in contrast to the build-up of long positions until 1926. The average deferment fee for all stocks in short-term clearing futures transactions was negative, indicating an accumulation of selling stocks. Ensuiko Sugar Refining (market capitalization ratio of 9.2\%), which was subject to short-term clearing futures transactions, was not immune to these effects because of its close relationship with Suzuki Shoten, a Zaibatsu Shosha (company syndicate) that went bankrupt during the Showa financial crisis. Ensuiko Sugar Refining's negative deferment fee increased temporarily. Although the stock market turmoil related to the financial moratorium was limited to the stocks of failed Suzuki Shoten-related companies and bank stocks, the degree of market efficiency bottomed out in the wake of the financial moratorium (turning point (a) in Table 6).

\begin{figure}[tbp]
 \caption{\small{Time-Varying Degree of Market Efficiency (TSE New Shares)}}
 \label{prewar_stock_fig4}
 \begin{center}
 \includegraphics[scale=0.4]{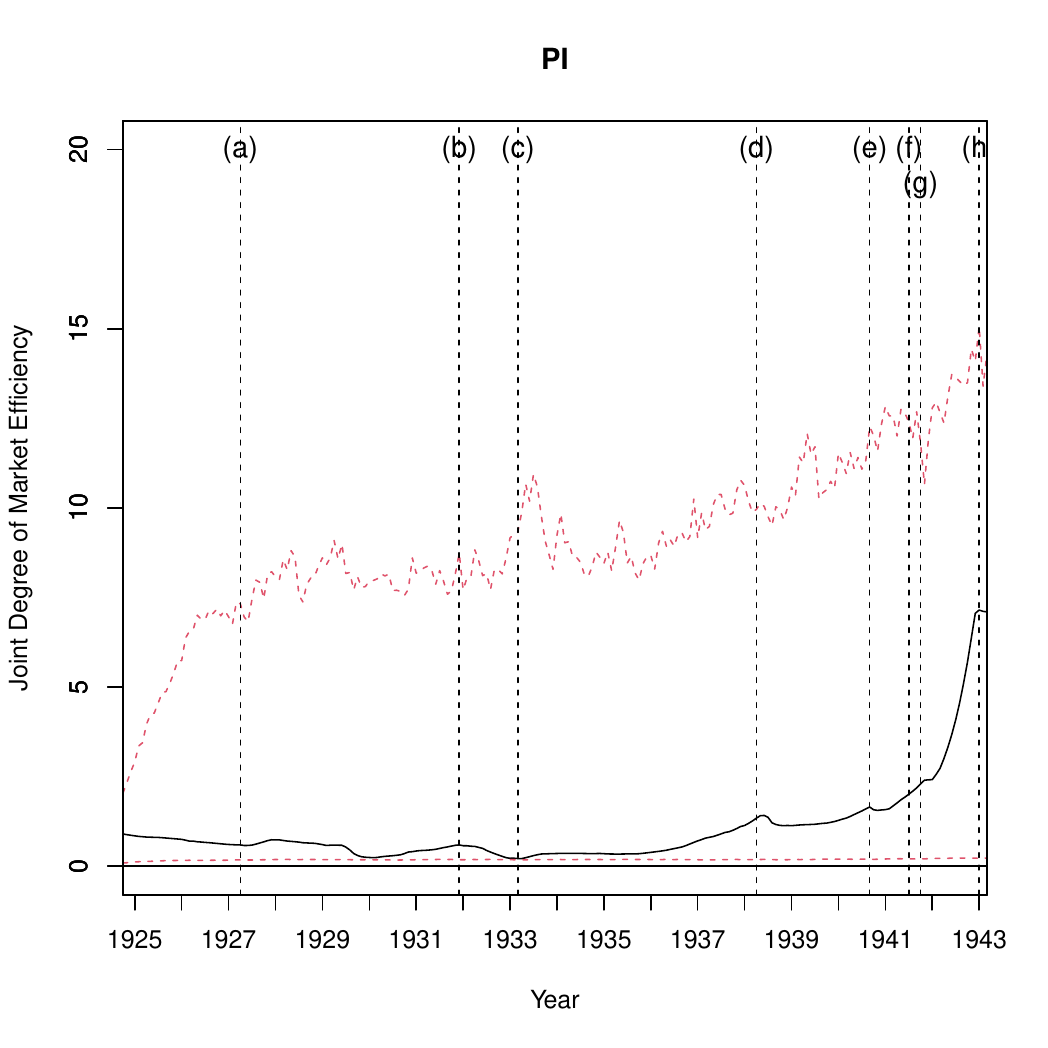}
 \includegraphics[scale=0.4]{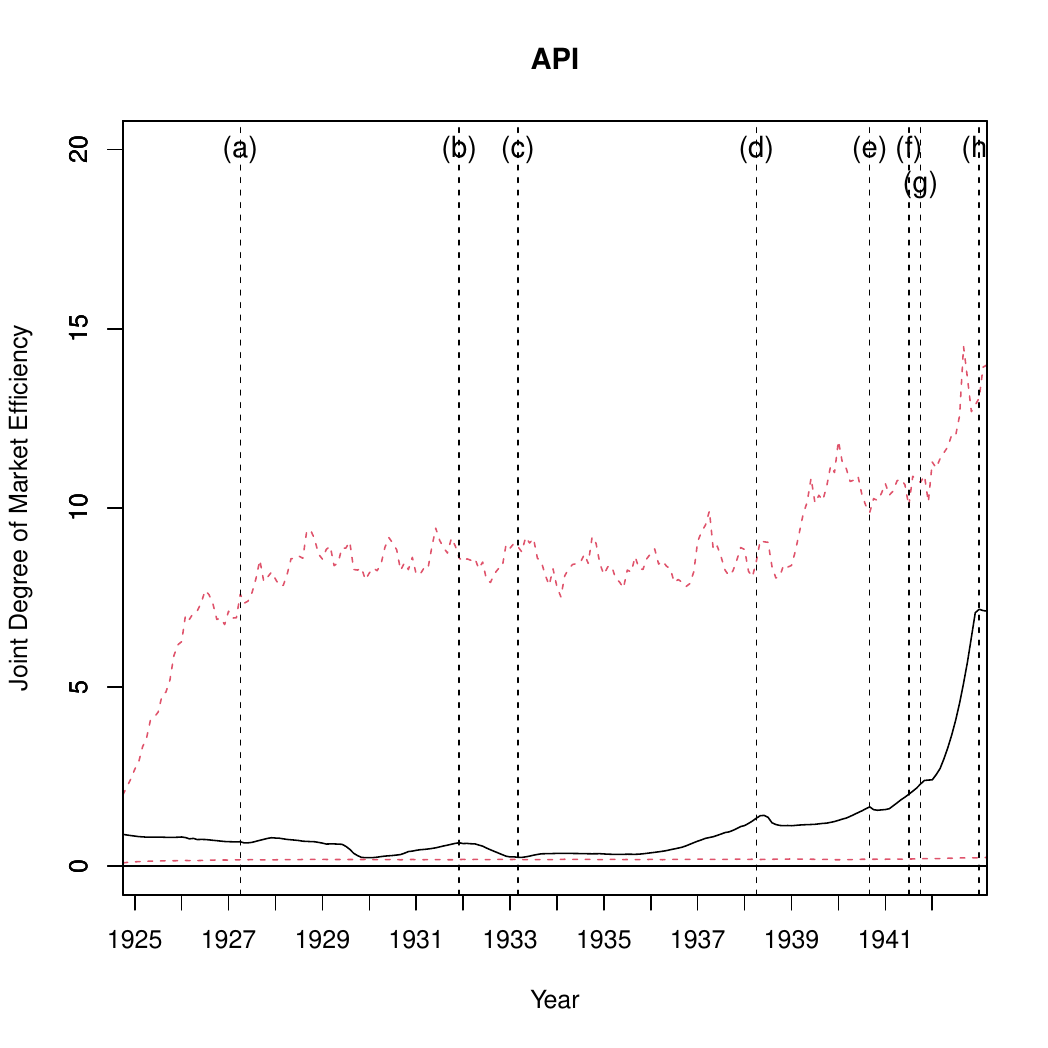}
 \includegraphics[scale=0.4]{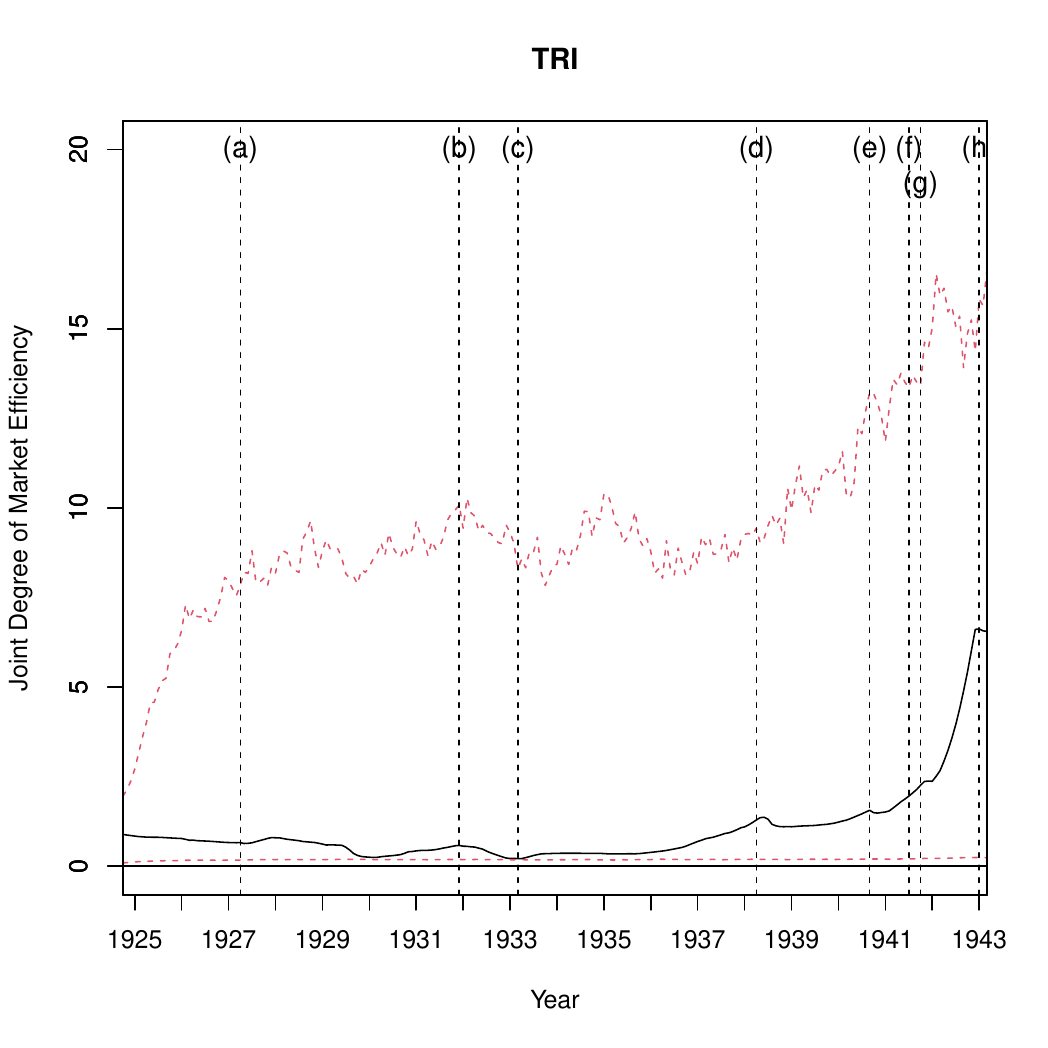}
\vspace*{3pt}
{
\begin{minipage}{400pt}\footnotesize
\underline{Note}: As for Figure \ref{prewar_stock_fig3}.
\end{minipage}}%
\end{center}
\end{figure}

Second, the {\it{Monthly Survey}} indicates that the negative factors related to lifting the ban increased and stock prices, especially the TSE's new shares, plummeted from April to May 1929, owing to the possibility of a changing exchange rate system. At the peak of this uncertainty, in July 1929, the Hamaguchi Cabinet issued a statement announcing the implementation of a gold embargo (the return to the gold standard was implemented in January 1930). This announcement clarified the outlook of market participants and raised the degree of market efficiency of the TSE's new shares, as depicted in Figure \ref{prewar_stock_fig4}. In the wake of the uncertain outcome, all negative factors were priced into the market to improve market efficiency.
\begin{center}
(Figure \ref{prewar_stock_fig4} around here)
\end{center}

Surprisingly, the Great Crash of October 1929 was not highlighted as a negative factor in the {\it{Monthly Survey}}, and market participants believed that the subsequent falling of stock prices until November 1930 was not significantly affected by overseas markets. The focus on commodity markets and trade as negative factors indicates that fixing the exchange rate to the old exchange rate, which was stronger than before, caused a sharp decline in commodity markets and highlighted concerns over the deterioration of the domestic economy owing to increasing imports, particularly from China.\footnote{As global commodity prices fell and the silver bullion began to decline, the exchange rate of Shanghai, which had adopted the silver standard, fell sharply, indicating a sharp increase in imports from China. Additionally, India, one of Japan's two largest export markets, increased its cotton cloth tariff applicable to Japanese goods in the early 1930s, worsening the trade situation.}

When the pound sterling fell after the U.K. announced the suspension of the gold standard system in September 1931, forecasts of the abandonment of the system intensified in Japan. Therefore, stock purchasing, in anticipation of a rebound in stock prices, accelerated toward the end of the year. The change in Japan's exchange rate policy significantly altered market participants' behaviors, along with a rapid increase in long positions.\footnote{See Bank of Japan, Research Department, {\it Monthly Survey} of September 1931.} The deferment fee for the TSE's new shares increased to 190 sen ($=1.90$ yen) in December 1931, the highest in the history for short-term clearing futures transactions.

Third, the degree of market efficiency based on the EQPI and TSE's new shares bottomed out in December 1931 with the ban on gold exports (turning point (b) in Table \ref{prewar_stock_table6}). This policy change cleared the negative uncertainty about the future and increased confidence in the prospects of market participants. In December 1931, Korekiyo Takahashi became Minister of Finance and decided to withdraw from the gold standard system. This caused a rapid appreciation of the U.S. dollar against the yen and an upward trend in the stock market.\footnote{The announcement of an additional payment of 12.50 yen in December 1931 was positively perceived by market participants, contributing to a sharp rise in the price of the TSE's new shares.} In July 1932, the Capital Flight Prevention Law was enacted; however, it did not directly prevent a decline in the exchange rate. Therefore, the yen continued to depreciate against the U.S. dollar until November 1932, when the Ministry of Finance ordinance was announced. According to \citet{fukai1941r70}, the Capital Flight Prevention Law did not restrict the freedom of transactions but left it to the self-restraint of exchange market participants.

The {\it{Monthly Survey}} reports that while overseas politics became a negative factor (international criticism of the Manchurian incident), the stock market was also affected by positive factors, such as the low exchange rate caused by the rapid depreciation of the yen, and by domestic finance and bonds, such as the underwriting of Japanese government bonds by the Bank of Japan.  

In November 1932, an ordinance of the Ministry of Finance required foreign exchange banks to report the contents of their foreign exchange transactions promptly. Apparently, the Japanese government's stance on exchange rates had shifted from a laissez-faire policy to one of preventing exchange rates from depreciating, as \citet{boj1948ffh} pointed out. The Foreign Exchange Control Law was promulgated in March 1933 and enforced in May 1933 (turning point (c) in Table \ref{prewar_stock_table6}). Uncertainty about the prospect of a major policy shift (i.e., ending the gold standard) had been resolved, and stock markets fully reflected this positive factor in prices, becoming more efficient. However, the elimination of uncertainty over the international financial system signaled the beginning of uncertainty in international relations, and the efficiency of the stock market peaked in the spring of 1933.

\subsection{Withdrawal of the U.S. Gold Standard and Strengthening of Control over the Japanese Economy (April 1933--August 1945)}
In April 1933, U.S. President Franklin D. Roosevelt announced the abolishment of the gold standard. This reversed the yen's depreciation in 1932 and instigated the depreciation of the U.S. dollar in 1933. Subsequently, the U.S. dollar and the pound remained stable against the yen until the beginning of WWI\hspace{-.1em}I. The increasing negative uncertainty of deteriorating international relations made it difficult for the stock market to factor in sufficient information. Our analysis is based on three chronological periods.

First, the trading ratio of long-term clearing futures transactions increased owing to the broad trading of shares, excluding the TSE's new shares. There were no significant changes in deferment fees. This finding suggests that trading was not conducted in an extraordinary market environment in which trading was biased in one direction (Figures \ref{prewar_stock_fig1} and \ref{prewar_stock_fig2}). Thereafter, according to the {\it{Monthly Survey}}, corporate performance affected the stock market in 10 of 48 factors in 1934 and 24 of 58 factors in 1935 (Table \ref{prewar_stock_table1}). While market participants recognized these items as a positive factor in 1934 owing to an increase in profits and dividends, they were perceived as a negative factor in 1935 because of a decrease in dividends. At the end of 1935, stocks were selected based on business performance, corresponding to a selective market that considered the merits and demerits of both businesses and companies.\footnote{See Bank of Japan, Research Department, {\it Monthly Survey} in December 1935.}

The so-called ``February 26 incident,'' an attempted coup d'\'{e}tat by army officers, ended Takahashi's economic policy. As military spending increased and government bonds were issued in large quantities, control over the government bond market strengthened, and government intervention in the market increased significantly.\footnote{In particular, the price fluctuations in the government bond market were restrained{\color{red},} and the government bond price support policy was strengthened (see \citet{hirayama2021jgb} for details).} Since the end of Takahashi's policy, market efficiency declined further amid the shift to a wartime regime and increased government intervention in financial markets, because of the increase in negative uncertainty. After 1937, the stock market became less volatile and the deferment fee for short-term clearing futures transactions stabilized, except in 1940 and 1941.\footnote{For the TSE's new shares, the deferment fee at times became negative owing to system problems and delisting issues.} The {\it{Monthly Survey}} suggests that the influence of the international political situation and domestic policies on stock price fluctuations increased after 1936. In addition to the deterioration of international relations with China and the Soviet Union, domestic policy conditions, such as stricter controls and higher taxes, became negative factors. Simultaneously, international conflicts in Europe and domestic policy issues, such as the establishment of stock price-keeping institutions and fiscal expansion, became positive factors for stock prices.

Although the TSE trading volume reached a prewar high in March 1937, the Second Sino--Japanese War broke out in July, and the value of the EQPI fluctuated widely until 1939. The {\it{Monthly Survey}} includes several descriptions of domestic and international policies. For the former, stock price-keeping operations (market intervention) by {\it{Dainippon Shoken Toshi}} and {\it{Seiho Shoken}} (stock price-keeping institutions) worked as positive factors, while tighter controls, such as the restriction of stock dividends by invoking Article 11 of the National Mobilization Law and the National Electricity Management Law, were negative factors. From then until the end of 1940, market efficiency improved slightly (turning point (d) in Table \ref{prewar_stock_table6}). For turning point (d), the outbreak of the Second Sino--Japanese War and concerns over its duration worked as negative factors, and the improvement of the war situation, such as the fall of Nanjing, was a positive factor (Table \ref{prewar_stock_table5}). 

Second, in September 1939, with expectations of an early end to the Second Sino--Japanese War and the declaration of war by the U.K. and France against Germany, the profits of export-related companies were predicted to expand and stock prices rose as a result. Meanwhile, the degree of market efficiency based on the EQPI increased slightly after the outbreak of the European War (see Figure \ref{prewar_stock_fig3}). 

In September 1940, the Tripartite Pact between Japan, Germany, and Italy was signed and the stock prices of the TSE's new shares fell by more than 14\% in a month. Subsequently, in December 1940, the Japanese government announced the Outline of the New Economic Order (turning point (e) in Table \ref{prewar_stock_table6}). This curtailed shareholder rights and accelerated the transfer of decision-making power to corporate managers and employees for wartime supplies. This negative factor disrupted the outlook of market participants, and the reaction mechanism of prices to information was paralyzed. Therefore, the market efficiency of the EQPI peaked. From then until July 1941, Japanese government intervention in the stock market became more extreme, and the degree of market efficiency plummeted (turning point (f) in Table \ref{prewar_stock_table6}). {\it{Nippon Shoken Toshi}} (one of the stock price-keeping institutions) was established to stabilize stock prices, and in March 1941, {\it{Nippon Kyodo Shoken}} (the other stock price-keeping institution) was established to bolster them. In July 1941, stock prices fell sharply because of the U.S. freeze on assets held by Japan. Consequently, deferment fees dropped sharply and the TSE's new shares recorded a low of 300 sen (negative $=3.00$ yen), while the lowest average of all short-term clearing futures transactions was 163 sen (negative $=1.63$ yen). Therefore, asset freezing significantly affected the Japanese stock market and caused a rapid increase in short positions. During this period, the government issued a loan order to the {\it{Industrial Bank of Japan}} to supply money and increase the stock prices. In August 1941, the Stock Price Control Ordinance was promulgated and enacted, which allowed the government to limit the minimum stock price.\footnote{The December 15, 1941 revision to this ordinance allowed the government to set a maximum price as well. However, the designation of the minimum and maximum price was not implemented until the end of WWI\hspace{-.1em}I. In fact, several stock price-keeping institutions intervened on their own in the stock market; \citet[pp. 232--35]{shibata2011mcw}.} At this juncture, peace negotiations seemed to have progressed. Therefore, market efficiency improved temporarily, cushioning the shock of the asset freeze. \citet{tse1940msr} stated that ``the stock price plummeted across the board due to a flood of buyers' sell-offs'' and that the stock market flexibly reflected negative factors in economic activity. The {\it Monthly Survey} reported that ``The government seems to have decided to give Nippon Kyodo Shoken a way to maintain all share prices without restrictions,'' indicating the cause of the subsequent rebound in stock prices. However, with the resignation of the Konoe Cabinet and the formation of the Tojo Cabinet, a ``war risk premium'' was imposed, and market efficiency deteriorated again (turning point (g) in Table \ref{prewar_stock_table6}). We believe that this confusion prevented market participants from fully pricing in negative information.

Third, when the Pacific War broke out after Japan attacked Pearl Harbor in December 1941, stock prices, particularly the TSE's new shares, increased significantly. The government intervened to restrain prices (selling intervention), because market volatility increased in response to rising expectations of victory. As described in \citet{shibata2011mcw}, {\it{Nippon Kyodo Shoken}} continued the selling intervention between mid-December 1941 and January 1942. The Stock Price Control Ordinance was revised to allow the establishment of maximum stock prices, and the Ministry of Finance imposed a capital gains tax (tax on trading profits of liquidation transactions) to restrain overheating transactions. Unlike the government bond market, in which the yield level was almost fixed, this market intervention was implemented to deal with excessive fluctuations from price elasticity in the stock market. In April 1942, {\it{Nippon Kyodo Shoken}} was merged into the newly formed {\it{Wartime Finance Bank}}, and the conventional stock price-keeping policy was changed to a stock price-stabilization policy after the outbreak of the war against the U.S., as \citet{fer1943arj} pointed out.

After the Battle of Midway, the Japanese government  attempted to restrain the rise in stock prices; however, once the stock market began to weaken, the government actively intervened to protect stock prices. During the war, the government implemented policies to control market volatility by preventing excessive increases and decreases in stock prices. After the Pacific War broke out in December 1941 until June 1943, the deferment fee remained stable, and there was little need for agencies to intentionally manipulate it. Moreover, market efficiency improved when many hoped they had a better chance of winning the war early by evaluating positive factors. Many early battles, such as the Battle of Midway, were reported in the Japanese media with an overestimation of the results of the war. However, after the fall of Guadalcanal in early 1943, it became impossible to hide the deterioration of the war situation. As the diluted war risk premium again emerged, market efficiency further deteriorated because of the increase in negative uncertainty (turning point (h) in Table \ref{prewar_stock_table6}). After 1943, market efficiency showed signs of deteriorating again. However, in the year following the outbreak of the Pacific War, market efficiency improved rapidly. This seems inconsistent with \citet{suzuki2012pwt}, who concludes that market efficiency deteriorated after the start of the Pacific War. However, as Figure \ref{prewar_stock_fig4} shows, the same results as those of \citet{suzuki2012pwt} are obtained when new TSE shares are used as stock returns.

A possible reason for this is that the government decided to delist the TSE's shares from stock exchanges in December 1942, and the various exchanges were reorganized into the Japan Stock Exchange in August 1943. Consequently, the deferment fee for the TSE's new shares decreased significantly. During this period, the trading ratio of short-term clearing futures transactions on the TSE declined significantly, whereas those of long-term clearing futures  and spot transactions increased. The degree of market efficiency of the TSE's new shares declined sharply with an increase in negative uncertainty. Whereas market efficiency of the EQPI improved until early 1943, market efficiency of the TSE's new shares deteriorated rapidly. As a result, when market efficiency is measured by the TSE's new shares, it declined significantly at the start of the Pacific War, while the EQPI which was less affected by these shares (less than 2\% of the market capitalization ratio) showed a divergence in efficiency until 1943, when the deteriorating war situation became apparent. As control of the exchange shifted from the Ministry of Commerce and Industry to the Ministry of Finance in December 1941, the prospect of delisting the TSE increased, and efficiency probably declined sharply over the year leading up to the announcement (Figure \ref{prewar_stock_fig4}). The market capitalization weight of the TSE's new shares was less than 2\%, according to the EQPI; hence, the effect of the degree of market efficiency on the EQPI was not significant.

\subsection{Summary of Changes in the Degree of Market Efficiency in the Early Showa Period}
Here, we summarize the historical interpretation from the perspective of the AMH as follows. First, we find that the relative change in market efficiency over time corresponds to major historical events, suggesting that the AMH is supported in the prewar and wartime Japanese stock market. For instance, the changes in deferment fees in Figure \ref{prewar_stock_fig2} show that market participants' trading tended to be skewed in one direction and insufficient trading takes place when large shocks occurred in the stock market. Thus, we consider that market efficiency deteriorated because of the inability to fully incorporate all information immediately (e.g., internal supply and demand, and changes in external political and economic conditions). Moreover, these trends to be more pronounced in the negative than in the positive for the stock market over the sample period. A possible reason is that the high impact and suddenness of such events (e.g., wars and disasters) make it more time-consuming to incorporate all information, reducing market efficiency. In practice, the market efficiency of the whole market improved in response to positive factors in the war situation, while the market efficiency of the TSE's new shares decreased in response to negative factors after the delisting in the period 1941--1943 (see Figures \ref{prewar_stock_fig3} and \ref{prewar_stock_fig4}).

Next, our historical interpretation suggests that the peculiarities of the prewar and wartime Japanese stock market affected efficiency. Our results show that the stock market was almost entirely efficient in an absolute sense throughout the sample period, which is inconsistent with the results of \citet{bassino2015iet}. We believe that the differences in the results can be attributed to differences in the datasets and econometric methodologies. In particular, their dataset does not include the WWI\hspace{-.1em}I period when volatility was controlled by the Stock Price Control Ordinance. Moreover, the fact that \citet{bassino2015iet} use the time-invariant GARCH-in-mean model as the verification method explains the difference from our results. In addition, the proportion of Japanese companies raising money through the stock market declined during the war, and the amount raised might have been low, as described in \citet{hoshi2001cfg}. However, a certain level of market efficiency might have been maintained over time owing to the existence of clearing transactions that were unaffected by the flow of funds. In practice, \citet{hirayama2020jep} confirms that the price elasticity of major stocks (Nippon Yusen Kaisha and Kanegafuchi Industries) was maintained in spot transactions even in 1945, when financial controls were further tightened after the period covered by his study. Furthermore, price controls in financial markets worked strongly for government bonds but had little impact on equity markets, which prevented them from over-rising and falling sharply. We conclude that market efficiency during the war was worse than in the early 1930s owing to government intervention but did not reach the level of inefficiency in an absolute sense.

\section{Concluding Remarks}\label{prewar_stock_sec7}
This study investigates the relationship between major historical events and the efficiency of the prewar and wartime Japanese stock market from the perspective of the AMH. In particular, we estimate the degree of market efficiency in the prewar and wartime Japanese stock market using \citeapos{ito2014ism}{ito2014ism,ito2016eme,ito2022aae} GLS-based time-varying parameter model, and employ primary historical documents to clarify the factors that caused efficiency to fluctuate over time from a historical perspective.

We summarize our empirical results as follows. First, the degree of market efficiency in the prewar and wartime Japanese stock market varied over time, corresponding to major historical events. This result supports the operation of the AMH in this market, in line with \citet{noda2016amh}, who test the AMH for the post-WWI\hspace{-.1em}I Japanese stock market. Second, the variation in market efficiency observed in this study differs significantly from that in previous studies, such as \citet{kataoka2004b} and \citet{bassino2015iet}. We find that this difference depends on whether the price index is capitalization weighted. Finally, stronger price controls after the mid-1930s and wartime market intervention by stock price maintenance agencies restrained price volatility. As Japanese government intervention in the market intensified throughout the 1930s, market efficiency declined as the war risk premium rose, especially from the time when the Pacific War became inevitable.



\section*{Acknowledgments}
The authors thank three anonymous referees, Brian Varian, the associate editor, Kohei Aono, Jean-Pascal Bassino, Carola Frydman, Eric Hilt, Ryoji Hiraguchi, Mikio Ito, Shinya Kajitani, Keiichi Morimoto, Daisuke Nagakura, Tatsuyoshi Okimoto, Kentaro Saito, Masato Shizume, Shiba Suzuki, Tatsuma Wada, Tomoaki Yamada, Takenobu Yuki, the conference participants at the Japan Society of Monetary Economics 2021 Autumn Meeting and the Western Economic Association International 95th Annual Conference, and the seminar participants at Keio University and Meiji University for their helpful comments and suggestions. Noda is grateful for the financial assistance provided by the Japan Society for the Promotion of Science Grant in Aid for Scientific Research (grant numbers 19K13747 and 23H00838) and the Japan Science and Technology Agency, Moonshot Research \& Development Program (grant number: JPMJMS2215). All data and computer codes used are available upon request.

\clearpage

\bigskip

\bigskip

\bigskip

\setcounter{table}{0}
\renewcommand{\thetable}{\arabic{table}}

\clearpage

\clearpage

\begin{landscape}
\begin{table}[tbp]
\caption{Factors Affecting Stock Prices Based on Bank of Japan's {\it{Monthly Survey}}}
\label{prewar_stock_table1}
\begin{center}
\resizebox{25cm}{!}{\begin{tabular}{ll|cccccccccccccccccc|c}\hline\hline
 &  & 1924 & 1925 & 1926 & 1927 & 1928 & 1929 & 1930 & 1931 & 1932 & 1933 & 1934 & 1935 & 1936 & 1937 & 1938 & 1939 & 1940 & 1941 & Total\\\hline
\multirow{8}*{Domestic} & Political Events & 1 & 2 & 2 & 2 & 7 & 8 & 3 & 3 & 5 & 6 & 8 & 3 & 3 & 7 & 3 & 1 & 1 & 1 & 66\\
 & Policy Matters & 4 & 2 & 5 & 3 & 6 & 8 & 9 & 2 & 9 & 7 & 9 & 4 & 18 & 18 & 17 & 12 & 16 & 5 & 154\\
 & Financial Markets & 2 & 9 & 9 & 18 & 8 & 3 & 1 & 8 & 14 & 10 & 5 & 1 & 4 & 10 & 3 & 6 & 4 & 1 & 116\\
 & Bond Markets & $-$ & 1 & $-$ & 1 & 4 & 2 & 1 & 7 & 3 & 3 & 2 & $-$ & 1 & 2 & $-$ & $-$ & $-$ & $-$ & 27\\
 & Commodity Markets & 5 & 7 & 11 & 5 & 8 & 3 & 11 & 2 & 9 & 6 & 2 & 7 & 4 & 8 & 4 & 4 & $-$ & 1 & 97\\
 & Foreign Exchange & $-$ & 1 & 4 & 3 & 1 & 7 & 1 & 4 & 11 & 12 & 2 & 2 & 2 & 1 & $-$ & $-$ & $-$ & $-$ & 51\\
 & Trade Status & 2 & 2 & 6 & 6 & 1 & $-$ & 11 & 8 & 5 & 4 & 2 & 3 & 5 & 4 & 7 & $-$ & 2 & 2 & 70\\
 & Corporate Performance & 3 & 11 & 7 & 8 & 6 & 5 & 7 & 6 & 9 & 7 & 10 & 24 & 17 & 5 & 8 & 7 & 4 & $-$ & 144\\\hline
\multirow{5}*{Overseas} & Political Events & 1 & 3 &  & 5 & 3 & 5 & 3 & 4 & 18 & 24 & 6 & 9 & 10 & 17 & 16 & 14 & 21 & 11 & 170\\
 & Economic Situation & 3 & $-$ & 1 & 1 & $-$ & 1 & $-$ & 3 & 5 & 4 & 1 & 4 & 1 & 1 & 3 & $-$ & $-$ & $-$ & 28\\
 & Financial Markets & $-$ & 3 & 1 & $-$ & $-$ & $-$ & 2 & 2 & 1 & 3 & $-$ & $-$ & $-$ & $-$ & $-$ & $-$ & $-$ & $-$ & 12\\
 & Commodity Markets & 1 & $-$ & 2 & $-$ & $-$ & $-$ & 1 & 8 & 1 & 5 & 1 & 1 & $-$ & 3 & $-$ & $-$ & $-$ & $-$ & 23\\
 & Stock Markets & 0 & 4 & 5 & 0 & 1 & 0 & 0 & 7 & 4 & 4 & 0 & 0 & 0 & 3 & 2 & 1 & 0 & 0 & 31\\\hline
Total &  & 22 & 45 & 53 & 52 & 45 & 42 & 50 & 64 & 94 & 95 & 48 & 58 & 65 & 79 & 63 & 45 & 48 & 21 & 989\\\hline\hline
\end{tabular}}
\vspace*{3pt}
\resizebox{25cm}{!}{\begin{minipage}{550pt}
\scriptsize
\underline{Source}: Bank of Japan, Research Department, {\it{Monthly Survey}} (June 1924--April 1941).
\end{minipage}}%
\end{center}
\end{table}
\end{landscape}

\clearpage

\clearpage

\begin{landscape}
\begin{table}[tbp]
\caption{Characteristics of Price Indexes in the Prewar and Wartime Japanese Stock Market}
\label{prewar_stock_table2}
\begin{center}
\begin{tabular}{l|lllcc}\hline\hline
Database & Type of Average Index & Sample Period & Frequency & Adjusted Index & Total Index\\\hline
Toyo Keizai Shimpo & Simple Arithmetic Average & 1916/01-44/05 & Monthly & No & No\\
Toyo Keizai Shimpo & Simple Arithmetic Average & 1930/06-40/12 & Daily & No & No\\
Bank of Japan & Simple Arithmetic Average & 1924/01-42/06 & Monthly & No & No\\
Tokyo Stock Exchange & Volume-weighted & 1921/01-45/08 & Monthly & No & No\\
Nippon Kangyo Bank & Simple Arithmetic Average & 1910/12-45/01 & Monthly & No & No\\
EQPI (\citet{hirayama2018erj}) & Capitalization-weighted & 1924/06-45/08 & Monthly & Yes & Yes\\\hline\hline
\end{tabular}
\vspace*{3pt}
{
\begin{minipage}{630pt}
\footnotesize
{\underline{Notes:}}
\begin{itemize}
\item[(1)] This table relies heavily on \citet{hirayama2018erj}.
\item[(2)] ``Adjusted Index'' denotes the modified price index in the case of ex-rights and additional paid-in capital.
\item[(3)] ``Total Index'' denotes the modified ``Adjusted Index,'' which additionally accounts for the dividend.
\end{itemize}
\end{minipage}}%
\end{center}
\end{table}
\end{landscape}

\clearpage

\begin{table}[tbp]
\caption{Descriptive Statistics and Unit Root Tests}
\label{prewar_stock_table3}
\begin{center}
\begin{tabular}{ccccccc}\hline\hline
 &  & $R_{PI}$ & $R_{API}$ & $R_{TRI}$ & $C_{DEF}$ & \\\hline
Mean &  & 0.0005  & 0.0009  & 0.0055  & $-0.0009$  & \\
SD &  & 0.0481  & 0.0474  & 0.0480  & 0.4009  & \\
Min &  & $-0.1633$  & $-0.1633$  & $-0.1580$  & $-3.5419$  & \\
Max &  & 0.1991  & 0.1991  & 0.1991  & 3.4484  & \\\hline
ADF-GLS &  & $-10.2868$  & $-10.2559$  & $-10.7104$  & $-15.1517$  & \\
Lags &  & 0  & 0  & 0  & 1  & \\
$\hat\phi$ &  & 0.3622  & 0.3648  & 0.3266  & $-0.2952$  & \\\hline
$\mathcal{N}$ &  & \multicolumn{4}{c}{228} & \\\hline\hline
\end{tabular}
\vspace*{3pt}
{
\begin{minipage}{430pt}
\footnotesize
{\underline{Notes:}}
\begin{itemize}
\item[(1)] ``$R_{PI}$,'' ``$R_{API}$,'' ``$R_{TRI}$,'' and ``$C_{DEF}$'' denote the returns on the price index, adjusted price index, and total return index; and changes in deferment fee, respectively.
\item[(2)] ``ADF-GLS'' denotes the ADF-GLS test statistics, ``Lags'' denotes the lag order selected by the MBIC, and ``$\hat\phi$'' denotes the coefficient vector in the GLS detrended series; see Equation (6) in \citet{ng2001lls}.
\item[(3)] In the ADF-GLS test, a model with a time trend and a constant is assumed. The critical value at the 1\% significance level for the ADF-GLS test is ``$-3.42$.''
\item[(4)] ``$\mathcal{N}$'' denotes the number of observations.
\item[(5)] R version 4.4.0 is used to compute the statistics.
\end{itemize}
\end{minipage}}%
\end{center}
\end{table}

\clearpage

\begin{table}[p]
\caption{Preliminary Estimations and Parameter Constancy Tests}
\label{prewar_stock_table4}
\begin{center}
\begin{tabular}{ccccccccccc}\hline\hline
 &  & $R_{PI,t}$ & $C_{DEF,t}$ &  & $R_{API,t}$ & $C_{DEF,t}$ &  & $R_{TRI,t}$ & $C_{DEF,t}$ & \\\cline{3-4}\cline{6-7}\cline{9-10}
\multirow{2}*{$Constant$} &  & 0.0001  & $-0.0029$  &  & 0.0004  & $-0.0030$  &  & 0.0044  & $-0.0039$  & \\
 &  & [0.0032]  & [0.0200]  &  & [0.0031]  & [0.0202]  &  & [0.0033]  & [0.0217]  & \\
\multirow{2}*{$R_{i,t-1}$} &  & 0.3463  & 1.6212  &  & 0.3431  & 1.5942  &  & 0.3067  & 1.5531  & \\
 &  & [0.0462]  & [0.6345]  &  & [0.0440]  & [0.6414]  &  & [0.0477]  & [0.6686]  & \\
\multirow{2}*{$C_{DEF,t-1}$} &  & 0.0004  & $-0.4743$  &  & 0.0003  & $-0.4726$  &  & 0.0007  & $-0.4742$  & \\
 &  & [0.0079]  & [0.1060]  &  & [0.0078]  & [0.1063]  &  & [0.0079]  & [0.1069]  & \\
\multirow{2}*{$R_{i,t-2}$} &  & $-0.1619$  & $-1.2968$  &  & $-0.1760$  & $-1.3005$  &  & $-0.1515$  & $-1.3392$  & \\
 &  & [0.0733]  & [0.3996]  &  & [0.0733]  & [0.4090]  &  & [0.0804]  & [0.3972]  & \\
\multirow{2}*{$C_{DEF,t-2}$} &  & $-0.0028$  & $-0.2659$  &  & $-0.0005$  & $-0.2637$  &  & $-0.0018$  & $-0.2592$  & \\
 &  & [0.0079]  & [0.1007]  &  & [0.0087]  & [0.1006]  &  & [0.0083]  & [0.0993]  & \\\hline
$\bar{R}^2$ &  & 0.0987  & 0.1922  &  & 0.0983  & 0.1903  &  & 0.0779  & 0.1921  & \\
$L_C$ &  & \multicolumn{2}{c}{23.0428} &  & \multicolumn{2}{c}{22.5818} &  & \multicolumn{2}{c}{22.6711} & \\\hline\hline
\end{tabular}
\vspace*{3pt}
{
\begin{minipage}{420pt}\footnotesize
{\underline{Notes:}}
 \begin{itemize}
  \item[(1)] ``$R_{i,t-p}$'' and ``$C_{DEF,t-p}$'' denote the VAR($p$) estimate for each stock return and changes in deferment fee.
  \item[(2)] ``$\bar{R}^2$'' and ``$L_C$'' denote adjusted $R^2$ and \citetapos{hansen1992a} joint $L$ statistic with variance.
  \item[(3)] \citetapos{newey1987sps} robust standard errors are given in brackets.
  \item[(4)] R version 4.4.0 is used to compute the estimates.
 \end{itemize}
\end{minipage}}%
\end{center}
\end{table}

\clearpage

\begin{table}[tbp]
\caption{Major Historical Events in Prewar and Watime Japan}
\label{prewar_stock_table5}
 \begin{center}
  \footnotesize{\begin{tabular}{clcp{10cm}c}\hline\hline
    & Period &  & Major historical event & \\\hline
    & Apr 1925 &  & The Bank of Japan reduced the official discount rate to 7.30\%. & \\
    & Oct 1926 &  & The Bank of Japan reduced the official discount rate to 6.57\%. & \\
    & Apr 1927 &  & Nationwide financial panic sparked when debates in the Diet revealed financial difficulties between the Bank of Taiwan and Suzuki \& Co. (Showa Financial Crisis). & \\
    & Apr--May 1927 &  & The Japanese government declared a moratorium on payments (bank holidays) on 22 April to last until 13 May. & \\
    & Jul 1929 &  & The Hamaguchi Cabinet (Junnosuke Inoue was appointed the Minister of Finance) announced that the gold embargo would be lifted. & \\
    & Oct 1929 &  & The Great Crash occurred. & \\\hline
    & Jan 1930 &  & The Japanese government lifted the gold embargo. & \\
    & Oct 1930 &  & The SEIHO SHOKEN (life insurance companies) started stock price-keeping operations. & \\
    & Sep 1931 &  & The United Kingdom abandoned the gold standard. & \\
    & Dec 1931 &  & The Inukai Cabinet (Korekiyo Takahashi was appointed the Minister of Finance) abandoned the gold standard. & \\
    & Nov 1932 &  & The announcement of the Ordinance of the Ministry of Finance clarified that the government's policy was to prevent the yen from falling. The Bank of Japan started underwriting government bonds. & \\
    & Feb 1933 &  & The prospect of Japan withdrawing from the League of Nations emerged. & \\
    & Mar 1933 &  & The Foreign Exchange Control Law was promulgated. & \\
    & Apr 1933 &  & The United States announced its abandonment of the gold standard. & \\
    & Feb 1936 &  & The February 26 incident occurred; Korekiyo Takahashi's economic policy ended. & \\
    & Mar 1937 &  & The Tokyo Stock Exchange recorded the highest trading volume in the prewar period. & \\
    & Jul 1937 &  & The Second Sino--Japanese War occurred. & \\
    & Apr 1938 &  & The National Mobilization Law was promulgated and enforced. & \\
    & Sep 1939 &  & World War I\hspace{-.1em}I broke out. & \\\hline
    & Jul 1941 &  & President Franklin Roosevelt froze all Japanese assets in the United States. & \\
    & Aug 1941 &  & The Stock Price Control Ordinance was promulgated and enforced. & \\
    & Dec 1941 &  & The Pacific War broke out. & \\
    & Jun 1942 &  & The Battle of Midway broke out. & \\
    & Dec 1942 &  & The Japanese government decided to delist Tokyo Stock Exchange shares. & \\
    & Aug 1943 &  & The Tokyo Stock Exchange delisted short-term clearing futures transactions. & \\
    & Jul 1944 &  & Operation Forager (the Battle of Saipan) ended. The Wartime Finance Bank started stock price-keeping operations. & \\
    & Mar 1945 &  & The Great Tokyo Air Raids occurred; the Wartime Finance Bank started stock price-keeping operations (unlimited stock purchases). In July, the Japan Stock Exchange started stock price-keeping operations. & \\
    & Aug 1945 &  & The Pacific War ended. & \\\hline\hline
 \end{tabular}}
 \vspace*{3pt}
 {
 \begin{minipage}{400pt}\footnotesize
  \underline{Note}: This table is mainly constructed following \citet{boj1993cfm}.
 \end{minipage}}%
 \end{center}
 \end{table}

\clearpage

\begin{landscape}
 \begin{table}[tbp]
  \caption{Turning Points in the Prewar and Wartime Japanese Stock Market from the Viewpoint of Market Efficiency}\label{prewar_stock_table6}
 \begin{center}
  \footnotesize{\begin{tabular}{cl|l|p{13cm}c}\hline\hline
    & Period & Degree of market efficiency & Major historical event & \\\hline
    & (a) Apr--May 1927 & Bottom Out & Nationwide financial panic sparked (Showa Financial Crisis). & \\
    & & & The Japanese government declared a moratorium on payments. & \\ 
    & (b) Dec 1931 & Start of the Rise & The Inukai Cabinet abandoned the gold standard. & \\
    & (c) Mar--Apr 1933 & Peak Out & Japan withdrew from the League of Nations. & \\
    & & & The Foreign Exchange Control Law was promulgated. & \\
    & & & The U.S. announced its abandonment of the gold standard. & \\
    & (d) Apr--May 1938 & Bottom Out & The National Mobilization Law was promulgated and enforced. & \\
    & (e) Sep--Dec 1940 & Peak Out & The Tripartite Pact between Japan, Germany and Italy was signed. & \\
    & & & The Japanese government announced an Outline of the New Economic Order. & \\
    & (f) Jul--Aug 1941 & Double Bottom (1) & President Franklin Roosevelt froze all Japanese assets in the U.S. & \\
    & & & The Stock Price Control Ordinance was promulgated and enforced. & \\
    & (g) Oct--Dec 1941 & Double Bottom (2) & The third Konoe Cabinet resigned en masse and the Tojo Cabinet was established. & \\
    & & & The Pacific War broke out. & \\
    & (h) Jan--Feb 1943 & Peak Out & The Japanese forces began to withdraw from Guadalcanal. & \\\hline\hline
 \end{tabular}}
 \end{center}
 \end{table}
\end{landscape}

\clearpage

\begin{thebibliography}{49}
\newcommand{\enquote}[1]{``#1''}
\expandafter\ifx\csname natexlab\endcsname\relax\def\natexlab#1{#1}\fi

\bibitem[{{Bank of Japan}(1948)}]{boj1948ffh}
{Bank of Japan} (1948), \enquote{Manshu Jihen Igo no Zaisei Kinyushi
  {\rm{[Fiscal and Financial History after the Manchurian Incident]}},} Tokyo,
  Japan.

\bibitem[{Barro(2006)}]{barro2006rda}
Barro, R.~J. (2006), \enquote{Rare Disasters and Asset Markets in the Twentieth
  Century,} \textit{QuarterJournal of Economics}, 121, 823--866.

\bibitem[{Bassino and Lagoarde-Segot(2015)}]{bassino2015iet}
Bassino, J. and Lagoarde-Segot, T. (2015), \enquote{Informational Efficiency in
  the Tokyo Stock Exchange, 1931--40,} \textit{Economic History Review}, 68,
  1226--1249.

\bibitem[{Berkman et~al.(2011)Berkman, Jacobsen, and Lee}]{berkman2011tvr}
Berkman, H., Jacobsen, B., and Lee, J. (2011), \enquote{Time-Varying Rare
  Disaster Risk and Stock Returns,} \textit{Journal of Financial Economics},
  101, 313--332.

\bibitem[{Caldara and Iacoviello(2022)}]{caldara2022mgr}
Caldara, D. and Iacoviello, M. (2022), \enquote{Measuring Geopolitical Risk,}
  \textit{American Economic Review}, 112, 1194--1225.

\bibitem[{Choi(1999)}]{choi1999trw}
Choi, I. (1999), \enquote{Testing the Random Walk Hypothesis for Real Exchange
  Rates,} \textit{Journal of Applied Econometrics}, 14, 293--308.

\bibitem[{Choudhry(2010)}]{choudhry2010ww2}
Choudhry, T. (2010), \enquote{World War I\hspace{-.1em}I Events and the Dow
  Jones Industrial Index,} \textit{Journal of Banking and Finance}, 34,
  1022--1031.

\bibitem[{Elliott et~al.(1996)Elliott, Rothenberg, and Stock}]{elliott1996eta}
Elliott, G., Rothenberg, T.~J., and Stock, J.~H. (1996), \enquote{Efficient
  Tests for an Autoregressive Unit Root,} \textit{Econometrica}, 64, 813--836.

\bibitem[{Engle et~al.(1987)Engle, Lilien, and Robins}]{engle1987etv}
Engle, R., Lilien, D., and Robins, R. (1987), \enquote{Estimating Time Varying
  Risk Premia in the Term Structure: The {ARCH}-M Model,}
  \textit{Econometrica}, 55, 391--407.

\bibitem[{Escanciano and Lobato(2009)}]{escanciano2009apt}
Escanciano, C. and Lobato, I. (2009), \enquote{An Automatic Portmanteau Test
  for Serial Correlation,} \textit{Journal of Econometrics}, 151, 140--149.

\bibitem[{Escanciano and Velasco(2006)}]{escanciano2006gst}
Escanciano, C. and Velasco, C. (2006), \enquote{Generalized Spectral Tests for
  the Martingale Difference Hypothesis,} \textit{Journal of Econometrics}, 134,
  151--185.

\bibitem[{Fama(1970)}]{fama1970ecm}
Fama, E.~F. (1970), \enquote{Efficient Capital Markets: A Review of Theory and
  Empirical Work,} \textit{Journal of Finance}, 25, 383--417.

\bibitem[{{Federation of Economic Research Institutes}(1943)}]{fer1943arj}
{Federation of Economic Research Institutes} (1943), \textit{Nihon Keizai
  Nenshi {\rm{[Annual Report on the Japanese Economy]}}}, Senshin Sha, Tokyo,
  Japan.

\bibitem[{Fisher(1921)}]{fisher1921bfi}
Fisher, I. (1921), \enquote{The Best Form of Index Number,} \textit{Quarterly
  Publications of the American Statistical Association}, 17, 533--537.

\bibitem[{Fujino and Akiyama(1977)}]{fujino1977sir}
Fujino, S. and Akiyama, R. (1977), \textit{Shoken to Rishiritsu: 1874--1975
  {\rm{[Securities and Interest Rates: 1874--1975]}}}, Hitotsubashi University
  Economic Research Institute, Japan Economic Statistics Literature Center.

\bibitem[{Fukai(1941)}]{fukai1941r70}
Fukai, E. (1941), \textit{Kaiko 70-nen \rm{[Reflections of 70 Years]}}, Iwanami
  Shoten, Tokyo.

\bibitem[{Hamao et~al.(2009)Hamao, Hoshi, and Okazaki}]{hamao2009lpd}
Hamao, Y., Hoshi, T., and Okazaki, T. (2009), \enquote{Listing Policy and
  Development of the Tokyo Stock Exchange in the Prewar Period,} in
  \textit{Financial Sector Development in the Pacific Rim, East Asia Seminar on
  Economics}, eds. Ito, T. and Rose, A., University of Chicago Press, vol.~18,
  pp. 51--87.

\bibitem[{Hansen(1992)}]{hansen1992a}
Hansen, B.~E. (1992), \enquote{Testing for Parameter Instability in Linear
  Models,} \textit{Journal of Policy Modeling}, 14, 517--533.

\bibitem[{Hirayama(2018)}]{hirayama2018erj}
Hirayama, K. (2018), \enquote{The Japanese Equity Performance Index in the
  Early Showa Era (in Japanese),} \textit{Journal of Financial and Securities
  Markets {\rm{(Japan Securities Research Institute)}}}, 101, 71--91.

\bibitem[{Hirayama(2020)}]{hirayama2020jep}
--- (2020), \enquote{The Japanese Equity Performance from 1944 to 1945 (in
  Japanese),} \textit{Journal of Financial and Securities Markets {{\rm (Japan
  Securities Research Institute)}}}, 109, 63--85.

\bibitem[{Hirayama(2021)}]{hirayama2021jgb}
--- (2021), \enquote{The Japanese Government Bond Performance from 1944 to 1945
  (in Japanese),} \textit{Journal of Financial and Securities Markets
  {\rm{(Japan Securities Research Institute)}}}, 113, 1--27.

\bibitem[{Hirayama(2022)}]{hirayama2022smf}
--- (2022), \enquote{Stock Market Functions and Short-term Clearing
  Transactions in the Prewar Period (in Japanese),} \textit{Journal of
  Financial and Securities Markets {\rm{(Japan Securities Research
  Institute)}}}, 117, 77--98.

\bibitem[{Hoshi and Kashyap(2001)}]{hoshi2001cfg}
Hoshi, T. and Kashyap, A. (2001), \textit{Corporate Finance and Governance in
  Japan: The Road to the Future}, MIT Press.

\bibitem[{{Institute for Monetary and Economic Studies, Bank of
  Japan}(1993)}]{boj1993cfm}
{Institute for Monetary and Economic Studies, Bank of Japan} (1993),
  \enquote{Chronology of Financial Matters in Japan {\rm{(in Japanese)}},}
  {Institute for Monetary and Economic Studies, Bank of Japan}, revised
  Edition.

\bibitem[{Ito et~al.(2014)Ito, Noda, and Wada}]{ito2014ism}
Ito, M., Noda, A., and Wada, T. (2014), \enquote{International Stock Market
  Efficiency: A Non-Bayesian Time-Varying Model Approach,} \textit{Applied
  Economics}, 46, 2744--2754.

\bibitem[{Ito et~al.(2016)Ito, Noda, and Wada}]{ito2016eme}
--- (2016), \enquote{The Evolution of Stock Market Efficiency in the US: A
  Non-Bayesian Time-Varying Model Approach,} \textit{Applied Economics}, 48,
  621--635.

\bibitem[{Ito et~al.(2022)Ito, Noda, and Wada}]{ito2022aae}
--- (2022), \enquote{An Alternative Estimation Method for Time-Varying
  Parameter Models,} \textit{Econometrics}, 10, 23.

\bibitem[{Ito and Sugiyama(2009)}]{ito2009mdt}
Ito, M. and Sugiyama, S. (2009), \enquote{Measuring the Degree of Time Varying
  Market Inefficiency,} \textit{Economics Letters}, 103, 62--64.

\bibitem[{Kataoka et~al.(2004)Kataoka, Maru, and Teranishi}]{kataoka2004b}
Kataoka, Y., Maru, J., and Teranishi, J. (2004), \enquote{An Analysis of Stock
  Market Efficiency in the Late Meiji Era, Part.2 {\rm{(in Japanese)}},}
  \textit{Journal of Financial and Securities Markets {\rm{(Japan Securities
  Research Institute)}}}, 48, 69--81.

\bibitem[{Kim et~al.(2011)Kim, Shamsuddin, and Lim}]{kim2011srp}
Kim, J.~H., Shamsuddin, A., and Lim, K.~P. (2011), \enquote{Stock Return
  Predictability and the Adaptive Markets Hypothesis: Evidence from
  Century-Long U.S. Data,} \textit{Journal of Empirical Finance}, 18, 868--879.

\bibitem[{Lim and Brooks(2011)}]{lim2011esm}
Lim, K.~P. and Brooks, R. (2011), \enquote{The Evolution of Stock Market
  Efficiency Over Time: A Survey of the Empirical Literature,} \textit{Journal
  of Economic Surveys}, 25, 69--108.

\bibitem[{Lim et~al.(2013)Lim, Luo, and Kim}]{lim2013aus}
Lim, K.~P., Luo, W., and Kim, J.~H. (2013), \enquote{Are {US} Stock Index
  Returns Predictable? Evidence from Automatic Autocorrelation-Based Tests,}
  \textit{Applied Economics}, 45, 953--962.

\bibitem[{Lo(2004)}]{lo2004amh}
Lo, A.~W. (2004), \enquote{The Adaptive Markets Hypothesis: Market Efficiency
  from an Evolutionary Perspective,} \textit{Journal of Portfolio Management},
  30, 15--29.

\bibitem[{Lo(2005)}]{lo2005rem}
--- (2005), \enquote{Reconciling Efficient Markets with Behavioral Finance: The
  Adaptive Markets Hypothesis,} \textit{Journal of Investment Consulting}, 7,
  21--44.

\bibitem[{Malkiel(2004)}]{malkiel2004rwd}
Malkiel, B.~G. (2004), \textit{A Random Walk Down Wall Street: The Time-Tested
  Strategy for Successful Investing}, W. W. Norton \& Company, Inc.

\bibitem[{Newey and West(1987)}]{newey1987sps}
Newey, W.~K. and West, K.~D. (1987), \enquote{A Simple, Positive Semi-Definite,
  Heteroskedasticity and Autocorrelation Consistent Covariance Matrix,}
  \textit{Econometrica}, 55, 703--708.

\bibitem[{Ng and Perron(2001)}]{ng2001lls}
Ng, S. and Perron, P. (2001), \enquote{Lag Length Selection and the
  Construction of Unit Root Tests with Good Size and Power,}
  \textit{Econometrica}, 69, 1519--1554.

\bibitem[{Noda(2016)}]{noda2016amh}
Noda, A. (2016), \enquote{A Test of the Adaptive Market Hypothesis using a
  Time-Varying AR Model in Japan,} \textit{Finance Research Letters}, 17,
  66--71.

\bibitem[{Okazaki(1999)}]{okazaki1999cg}
Okazaki, T. (1999), \enquote{Corporate Governance,} in \textit{The Japanese
  Economic System and its Historical Origins}, eds. Okuno-Fujiwara, M. and
  Okazaki, T., Oxford University Press.

\bibitem[{Reitz(1988)}]{reitz1988erp}
Reitz, T.~A. (1988), \enquote{The Equity Risk Premium: A Solution,}
  \textit{Journal of Monetary Economics}, 22, 117--131.

\bibitem[{Roberts(1967)}]{roberts1967scp}
Roberts, H. (1967), \enquote{Statistical versus Clinical Prediction of the
  Stock Market,} Center for Research in Security Prices, University of Chicago.

\bibitem[{Saito(2016)}]{saito2016pps}
Saito, N. (2016), \enquote{Part-Paid Stock System in Prewar Japan: New
  Perspectives for Historical Analysis {\rm{(in Japanese)}},} \textit{Journal
  of Global and Inter-Cultural Studies}, 18, 81--102.

\bibitem[{Schwarz(1978)}]{schwarz1978edm}
Schwarz, G. (1978), \enquote{Estimating the Dimension of a Model,}
  \textit{Annals of Statistics}, 6, 461--464.

\bibitem[{Shibata(2011)}]{shibata2011mcw}
Shibata, Y. (2011), \textit{Senji Nihon no Kinyu Tosei {\rm{[Monetary Controls
  in Wartime Japan]}}}, Nihon Keizai Hyoron Sha, Tokyo.

\bibitem[{Shiller(2005)}]{shiller2005ies}
Shiller, R.~J. (2005), \textit{Irrational Exuberance: Second Edition},
  Princeton University Press.

\bibitem[{Suzuki(2012)}]{suzuki2012pwt}
Suzuki, S. (2012), \enquote{Pacific War and Tokyo Stock Exchange Daily
  Data:1941-1943 {\rm{(in Japanese)}},} \textit{Bulletin of Economic Studies
  {\rm{(Meisei University)}}}, 44, 39--51.

\bibitem[{{Tokyo Stock Exchange}(1938)}]{tse1938htse}
{Tokyo Stock Exchange} (1938), \textit{History of the Tokyo Stock Exchange
  \rm{(in Japanese)}}, vol.~3, Tokyo Stock Exchange.

\bibitem[{{Tokyo Stock Exchange}(1940)}]{tse1940msr}
--- (1940), \textit{Monthly Statistics Report \rm{(in Japanese), July 1940}},
  Tokyo Stock Exchange.

\bibitem[{Verdickt(2020)}]{verdickt2020ewr}
Verdickt, G. (2020), \enquote{The Effect of War Risk on Managerial and Investor
  Behavior: Evidence from the Brussels Stock Exchange in the pre-1914 Era,}
  \textit{Journal of Economic History}, 80, 629--669.

\end{thebibliography}
\end{document}